\documentclass[12pt, aps,prl, reprint,superscriptaddress, longbibliography, noeprint]{revtex4-1}
\usepackage{graphicx}
\usepackage{amsmath}
\usepackage{amssymb}
\usepackage[version=4]{mhchem} 
\usepackage[per-mode = symbol]{siunitx} 
\usepackage{gensymb}	
\usepackage{hyperref}
\hypersetup{colorlinks=true, linkcolor=blue, citecolor=red, urlcolor=blue} 
\usepackage[T1]{fontenc}	
\usepackage{placeins}
\usepackage{mathtools}
\usepackage{xcolor}
\usepackage{natbib}
\usepackage{dcolumn}
\usepackage{bm}
\usepackage{tabularx}
\usepackage{multirow}
\usepackage{array}
\usepackage{float}
\usepackage[english]{babel}
\usepackage[normalem]{ulem}
\usepackage{enumerate}

\setcitestyle{super}

\AtBeginDocument{%
	\newwrite\bibnotes
	\def\bibnotesext{Notes.bib}
	\immediate\openout\bibnotes=\jobname\bibnotesext
	\immediate\write\bibnotes{@CONTROL{REVTEX41Control}}
	\immediate\write\bibnotes{@CONTROL{%
			apsrev41Control,author="08",editor="1",pages="1",title="0",year="1"}}
	\if@filesw
	\immediate\write\@auxout{\string\citation{apsrev41Control}}%
	\fi
}%

\begin{document}

\begin{abstract}
Strain engineering has been extended recently to the ultrafast timescales, driving metal-insulator phase transitions and the propagation of ultrasonic demagnetization fronts. However, the non-linear lattice dynamics underpinning interfacial optoelectronic phase switching have not yet been addressed. Here we focus on the lattice dynamics initiated by impulsive resonant excitation of polar lattice vibrations in \ce{LaAlO3} single crystals, one of the most widely utilized substrates for oxide electronics. We show that ionic Raman scattering drives coherent oxygen octahedra rotations around a high-symmetry crystal axis and we identify, by means of DFT calculations, the underlying phonon-phonon coupling channel. Resonant lattice excitation is shown to generate longitudinal and transverse acoustic wavepackets, enabled by anisotropic optically-induced strain in and out of equilibrium. Importantly, shear strain wavepackets are found to be generated with extraordinary efficiency at the phonon resonance, being comparable in amplitude to the more conventional longitudinal acoustic waves, opening exciting perspectives for ultrafast material control.
\end{abstract}

\title{Tunable shear strain from resonantly driven optical phonons}

\date{\today} 

\author{J.R.~Hortensius}
\thanks{equal contribution}
\affiliation{Kavli Institute of Nanoscience,
	Delft University of Technology, P.O. Box 5046, 2600 GA Delft, The Netherlands}
\email{j.r.hortensius@tudelft.nl}
\author{D.~Afanasiev}
\thanks{equal contribution}
\affiliation{Kavli Institute of Nanoscience,
	Delft University of Technology, P.O. Box 5046, 2600 GA Delft, The Netherlands}
\author{A.~Sasani}
\affiliation{CESAM QMAT Physique Th\'eorique des Mat\'eriaux,
	Universit\'e de Li\`ege, B-4000 Sart Tilman, Belgium}
\author{E.~Bousquet}
\affiliation{CESAM QMAT Physique Th\'eorique des Mat\'eriaux,
	Universit\'e de Li\`ege, B-4000 Sart Tilman, Belgium}
\author{A.~D.~Caviglia}
\affiliation{Kavli Institute of Nanoscience,
	Delft University of Technology, P.O. Box 5046, 2600 GA Delft, The Netherlands}
\email{Contact email: a.caviglia@tudelft.nl}

\maketitle
Epitaxy can be used to impose misfit strain capable of altering the properties of materials. Notable examples include the enhancement of ferroelectric and ferromagnetic order\cite{artc:Schlom2014} and even the engineering of artificial multiferroics at room temperature\cite{Mundy2016}. Whereas static strain engineering is a well-established paradigm\cite{artc:Haeni2004,artc:Schlom2007, artc:Sando2013}, ultrafast strain engineering has emerged only recently as an effective method to drive optoelectronic phase switching\cite{artc:Caviglia2012}. In this approach, ultrashort pulses of light are tuned in resonance with an infrared-active atomic vibration of a substrate, in order to transform the structural and electronic properties of an epitaxial thin film. This mechanism, applied extensively to insulating lanthanum aluminate (\ce{LaAlO3}) substrates, governs metal-insulator transitions \cite{artc:Caviglia2012}, ultrasonic magnetic dynamics\cite{artc:Forst2015}, and sonic lattice waves \cite{artc:Forst2017} in strongly correlated thin films. However, the nature of the non-linear lattice dynamics initiated in the substrate material is not yet fully understood. Here we show that resonant impulsive excitation of a polar Al-O stretching of the \ce{LaAlO3} crystal lattice drives non-polar coherent rotations of oxygen octahedra via ionic Raman scattering.  Moreover, the anisotropic optically-induced stress generates propagating longitudinal and transverse acoustic wavepackets. Importantly, shear strain wavepackets are found to be produced with extraordinary efficiency at the phonon resonance, being of comparable amplitude to the more conventional longitudinal acoustic waves. These results uncover an hitherto unknown microscopic feature of ultrafast strain engineering that opens new perspectives for material control via tunable shear strain.

\begin{figure}
	\centering
	\includegraphics[width = \linewidth]{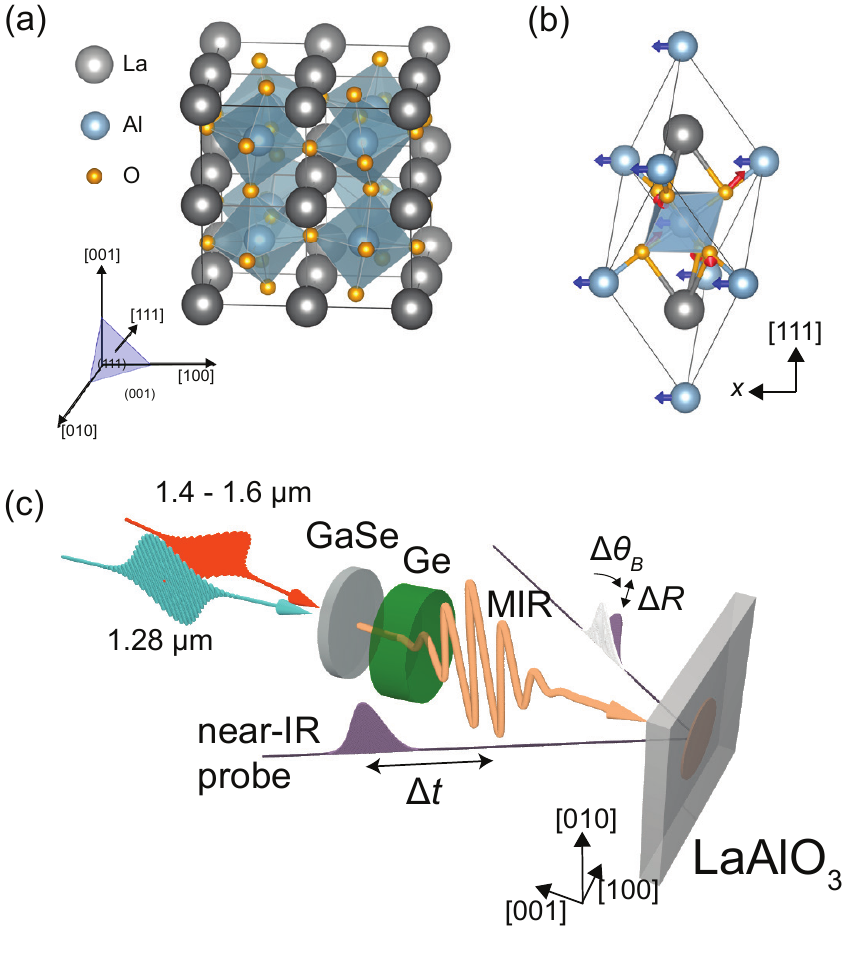}
	\caption{ (a) The crystal structure of \ce{LaAlO3}, with indications of the different crystallographic directions. The rhombohedral distortion from the high-temperature cubic phase is due to out-of-phase rotations of the oxygen octahedra about the $[111]\textsubscript{pc}$ axis, schematically depicted by a blue arrow. (b) Atomic motion corresponding to the infrared-active E\textsubscript{u} stretching mode, polarized in the (111) plane in the $x$-direction. (c) Schematic illustration of the experimental scheme. The mid-IR pulse is generated by difference frequency mixing from two near-infrared pulses in a GaSe crystal, after which the mid-IR pulses are filtered by a germanium (Ge) filter. Following the mid-IR excitation, the ensuing changes in optical properties are probed with a time-delayed near-infrared pulse. The pump-induced changes to the reflection intensity $\Delta R$ and rotation $\Delta \theta_\textrm{B}$ of the polarization plane are monitored.}
	\label{Fig:Figure1}
\end{figure}

\begin{figure}[ht!]
	\centering
	\includegraphics[width = \linewidth]{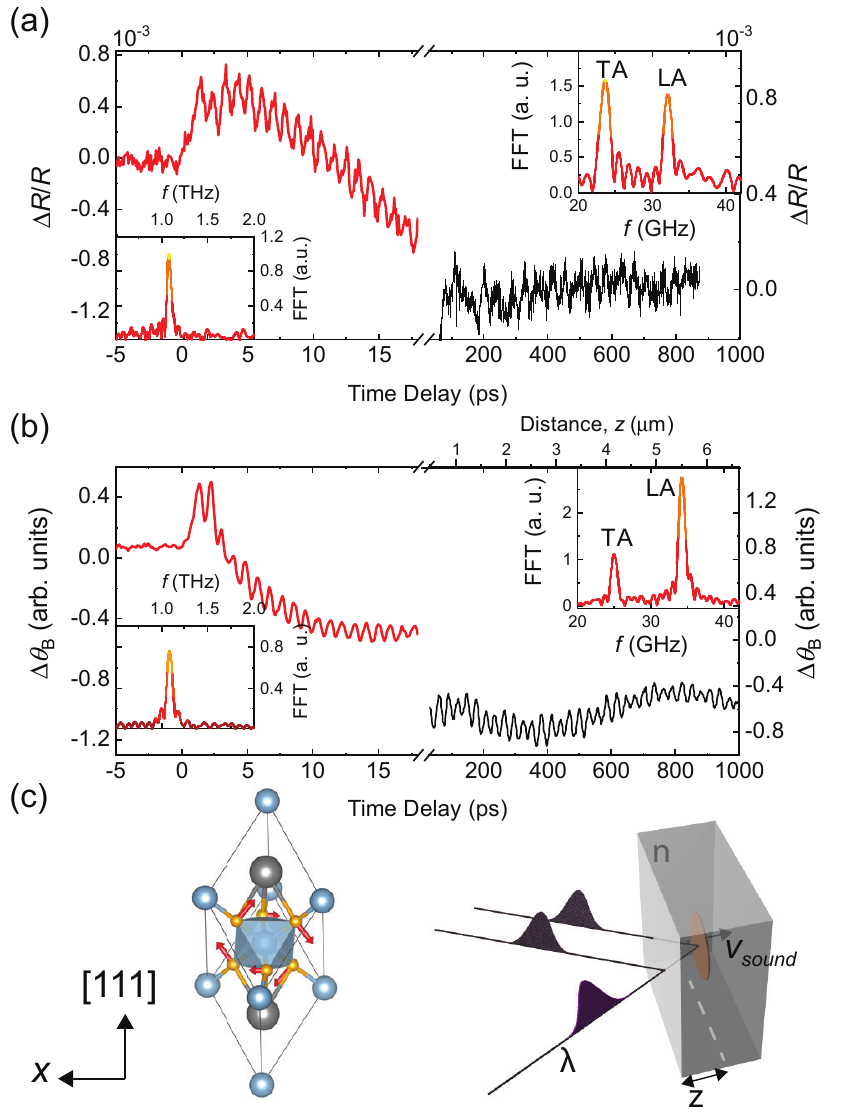}
	\caption{Transient changes in the intensity of the reflected \SI{1.5}{\electronvolt} probe pulse $\Delta R/R$ (a) and the polarization rotation $\Delta\theta$\textsubscript{B} of the probe pulse (b) after excitation of the \ce{LaAlO3} substrate with a mid-infrared pump pulse of \SI{86}{\milli\electronvolt} (a) and \SI{89}{\milli\electronvolt} (b) and a fluence of \SI{10}{\milli\joule\per\centi\meter\squared}. The insets show the Fourier spectra of the signal. The top axis in (b) shows the distance $z$ which the longitudinal sound wave has propagated at that
	time. (c) The real-space atomic motion corresponding to the excited E\textsubscript{g} mode indicated with arrows (left) and a schematic picture of a strain wave propagating with speed $v_{\text{sound}}$ leading to interference between probe light (wavelength $\lambda$) reflected at the interface and scattered at the strain wave, depending on the distance $z$ which the strain wave has propagated. (right)}
	\label{Fig:Figure2}
\end{figure}

We investigate light-induced structural dynamics in \ce{LaAlO3}, an insulating substrate utilized extensively in oxide electronics for the epitaxy of correlated materials, including high-T$_\textrm{c}$ cuprate superconductors\cite{artc:Wordenweber1999}, magnetoresistive manganites\cite{artc:Prellier2001} and nickelates\cite{artc:Middey2016}. At room temperature, \ce{LaAlO3} exhibits a distorted perovskite structure (rhombohedral space group \textit{R$\bar{3}$c}, see Fig.~\ref{Fig:Figure1}a). In our experiments we use commercially available \SI[product-units = single]{5x5}{\milli\meter} (001) \ce{LaAlO3} single crystals with a thickness of \SI{0.5}{\milli\meter}. To drive the lattice vibrations resonantly, we use ultrashort (\SI{200}{\femto\second}) pump pulses in the mid-IR frequency range. The pulses are carrier-envelope-stable (CEP-stable) and are generated in a \SI{0.35}{\milli\meter} thick \ce{GaSe} crystal by difference frequency mixing the output of two optical parametric amplifiers (OPAs). The OPAs share the same white light, generated in a sapphire crystal, by the output of a laser amplifier (800 nm, 100 fs, 5 mJ, 1 kHz), which ensures CEP-stability of the pulses\cite{artc:Sell2008}. We tune the photon energy of the pump pulses continuously across the highest-frequency E$_\textrm{u}$ phonon resonance~\cite{artc:Zhang1994,artc:Abrashev1999}, schematically depicted in Fig.1b. The tuning of the pump photon energy was done in the range \SIrange[range-units = single, range-phrase = --]{70}{180}{\milli\electronvolt} and allowed us to compare dynamics excited in the optical transparency window, with structural transient dynamics induced by pulses tuned in resonance with the lattice vibrational mode. 
The wide bandgap of \ce{LaAlO3} (\SI{5.6}{\electronvolt}\cite{artc:Lim2002}), as well as the absence of electronic in-gap states\cite{artc:Chernova2017}, ensures the purely structural nature of the photoinduced response.

Time-resolved optical reflectivity and birefringence measurements are performed using near infrared probe pulses (\SI{100}{\femto\second}, \SI{1.5}{\electronvolt}, 1 kHz), in two complementary experimental geometries, schematically illustrated in Fig.~\ref{Fig:Figure1}c. In the first scheme we monitor the transient differential reflectivity $\Delta R$ using a balanced photodetector. The structural dynamics initiated by the pump pulse modulate the sample's dielectric function resulting in a perturbation of the refractive index $n$, being directly imprinted on the $\Delta R$ signal. In the second scheme we track the transient optical birefringence $\Delta \theta_\textnormal{B}$ using an optical polarization bridge (Wollaston prism) and a balanced photodetector. The elementary vibrations of \ce{LaAlO3} are intrinsically highly anisotropic and coherent dynamics of these modes can modify the off-diagonal components of the permittivity tensor, thereby resulting in a transient birefringence. 
In both experimental configurations, the probe pulses are focused to a spot with a diameter of \SI{80}{\micro\meter}. The spatial overlap between the pump and probe pulses is obtained by co-propagation of the beams, using an off-axis parabolic mirror, which focuses the pump beam to a spot of about \SI{150}{\micro\meter}.

Measurements of transient changes to both the reflectivity and birefringence, using pump pulses at the photon energy tuned in resonance with the E$_u$ phonon mode, $h\nu\simeq$ 85 meV, reveal multiple oscillatory responses at frequencies well below the one of the pump, see Fig.~\ref{Fig:Figure2}a,b. The highest-frequency oscillation is centered at \SI{1.1}{\tera\hertz} and is assigned to the Raman-active E$_\textrm{g}$ soft mode of \ce{LaAlO3}~\cite{artc:Scott1969, artc:Abrashev1999} associated with a rhombohedral instability of the \textit{R$\bar{3}$c} lattice structure. This mode is comprised of rotations of the oxygen octahedra around an axis perpendicular to the [111] pseudocubic direction as shown in Fig.~\ref{Fig:Figure2}c. 
The longer time delay further reveals oscillatory components at two discrete frequencies $f_{\textnormal{TA}}$ and $f_{\textnormal{LA}}$ in the GHz frequency range. This pattern originates from interference between light pulses reflected at the crystal surface and reflections from an acoustic wavefront propagating into the bulk (Fig.~\ref{Fig:Figure2}c). The frequency of the oscillations $f$ is related to the refractive index $n$ of the material at the probe wavelength, the speed of sound $v_\textnormal{s}$, the angle $\theta$ w.r.t. the sample normal and the wavelength $\lambda$ of the probe by the relation \cite{artc:Matsuda2004} $f = 2nv_\textnormal{s} \cos(\theta)/\lambda$.
In our experiments we vary the angle of incidence of the probe beam and find that, while the frequency of the Raman oscillation remains unchanged, the frequency of the GHz oscillations decreases in agreement with the relation shown above. We extract the corresponding propagation velocities, obtaining $v_{\textnormal{LA}} = \SI{6.67}{\kilo\meter\per\second}$ and  $v_{\textnormal{TA}} = \SI{4.87}{\kilo\meter\per\second}$ matching well the speed of longitudinal acoustic (LA) and transverse (TA) phonons in \ce{LaAlO3} propagating along the [001] direction~\cite{artc:Carpenter2010}. Therefore our experiments show that optical excitation with ultrashort resonant mid-IR pulses initiates coherent structural dynamics in both the acoustic and optical branches of the phonon spectrum. 
Although the optical excitation of a longitudinal acoustic wavefront is expected from electrostriction in \ce{LaAlO3} \cite{artc:Cancellieri2011} and/or optical absorption\cite{artc:Ruello2015}, the optical generation of shear strain requires the presence of an equilibrium or light-induced structural anisotropy. We discuss this aspect below, after the analysis of the THz Raman mode.

\begin{figure}
	\centering
	\includegraphics[width = \linewidth]{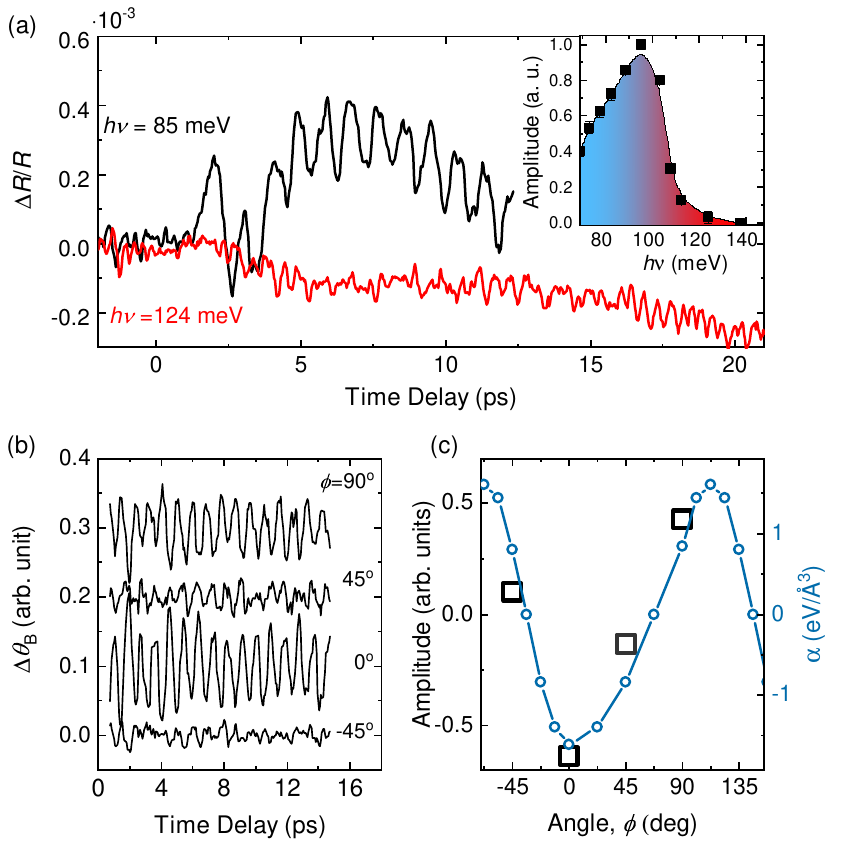}
	\caption{(a) Time resolved reflectivity changes of the probe pulse at the central energy of \SI{1.5}{\electronvolt} after mid-IR excitation at photon energies of \SI{83}{\milli\electronvolt} and \SI{124}{\milli\electronvolt}. Inset: the amplitude of the coherent phonon oscillations at \SI{1.1}{\tera\hertz}, as extracted from a sine fit to the transient reflectivity changes for different pump photon energies. The solid line serves as a guide to the eye. 
	(b) Time-resolved measurements of the transient polarization rotation of the probe pulse after excitation with the pump pulses having a polarization of $\phi$ degrees with respect to the pseudocubic axis at a pump energy of \SI{89}{\milli\electronvolt}. (c) Amplitude of the observed E\textsubscript{g} mode excitation (black filled squares) and calculated coupling of the excited phonons to the E$_g$ mode (blue circles) vs pump polarization angle $\phi$ with a sine fit.}
	\label{Fig:Figure3}
\end{figure}

To unveil the mechanism of excitation of the Raman-active mode, we vary the pump photon energy across the E$_\textrm{u}$ phonon resonance. Fig.~\ref{Fig:Figure3}a compares time-resolved transient reflectivity $\Delta R/R$ induced by pumping at the phonon resonance (\SI{85}{\milli\electronvolt}) with off-resonant pumping (\SI{124}{\milli\electronvolt}), revealing a striking selectivity of the low-energy mode excitation. The inset in Figure~\ref{Fig:Figure3}a shows that the amplitude of the excited Raman E$_\textrm{g}$ mode measured for various photon energies increases strongly to a peak at \SI{95}{\milli\electronvolt} in vicinity of the absorption peak attributed to the E$_\textrm{u}$ phonon mode. 
Recently, ionic Raman scattering (IRS) or nonlinear phononics, was proposed as a mechanism for resonant activation of coherent low-energy Raman-active (non-polar) phonon modes upon pumping the absorption lines of infrared-active (polar) lattice vibrations~\cite{artc:Forst2011_2}. 
The mechanism of the coupling between R- and IR-active modes can be described by introducing an invariant non-linear term $\alpha Q_{\text{IR}}^2 Q_{\text{R}}$ in the lattice potential, with $\alpha$ defining the strength of the coupling and $Q$ corresponding to a normal coordinate of a phonon mode. Because the symmetry representation E$_g$ of the phonon mode transforms as E$_u\otimes$E$_u$ under the symmetry operations of the rhombohedral phase of \ce{LaAlO3}, the $Q_{\text{E}_u}^2 Q_{\text{E}_g}$ coupling term is symmetry-allowed. To verify these mechanisms, we perform density functional theory (DFT) calculations with the ABINIT code\cite{artc:gonze2020} to fit a non-linear phonon-phonon model potential of bulk $R\bar{3}c$ \ce{LaAlO3} (see supplemental material section II).

In Fig.~\ref{Fig:Figure3}c we show the evolution of the coupling coefficient $\alpha$ with respect to the pump polarization angle $\phi$, such that $\phi$=0 corresponds to the pump polarization oriented along the (100) axis. It evolves as a periodic function with extrema around $\phi$= 0$^\circ$ and $112.5^\circ$. Such a non-trivial periodicity is a consequence of the projection of the experimental pseudo cubic reference frame to the natural rhombohedral one containing a high-symmetry 3-fold rotation axis along the pseudocubic [111] direction, see supplemental material, II.C.  
To verify this behavior, we measured the amplitude of the E$_\textrm{g}$ oscillation for the pump polarizations oriented along several pseudocubic crystallographic directions, see Fig.~\ref{Fig:Figure3}b.
Figure~\ref{Fig:Figure3}c summarizes the observations showing a good agreement with predictions of the non-linear phonon model built from DFT. This confirms that excitation of the E$_g$ Raman-active mode is governed by the IRS mechanism. Relevance of this mechanism is further corroborated by measurements of the fluence dependence of the Raman mode amplitude, revealing a linear increase in the amplitude. This is the first experimental observation of nonlinear phononics in a wide bandgap insulator in conditions allowing exclusively coherent phonon-phonon coupling. In this sense IRS differs substantially from regular impulsive stimulated Raman scattering (ISRS) in which excitation of coherent phonons is mediated by virtual electronic transitions.

\begin{figure}
	\centering
	\includegraphics[width = \linewidth]{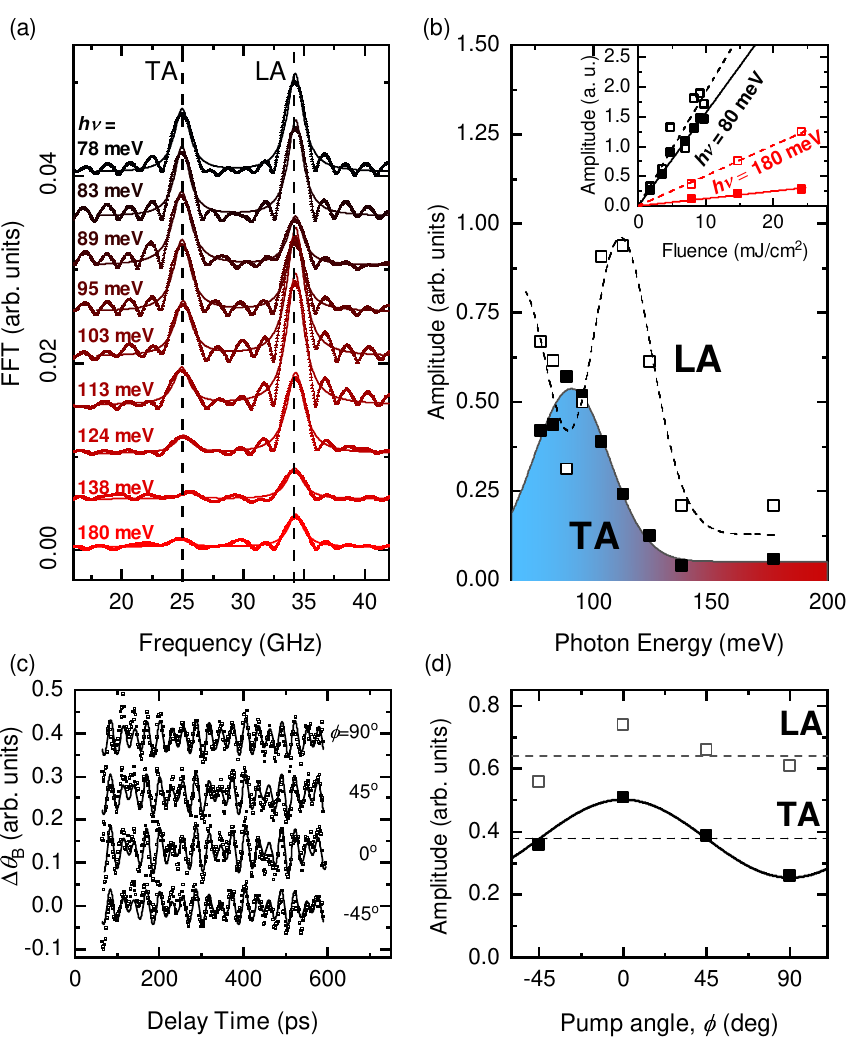}
	\caption{ (a) Fourier transforms of the measured polarization rotation signals corresponding to strain wave propagation after excitation with different pump energies. (b) Pump wavelength dependent amplitude of the two oscillations corresponding to LA and TA strain waves for fixed pump and probe polarizations as obtained from the Fourier transforms shown in panel a. The solid and dashed lines are guides to the eyes. The inset shows the amplitude of the LA (open markers) and TA (filled markers) oscillations vs pump fluence for resonant and non-resonant excitation. (c) Time-resolved measurements of the transient polarization rotation of the probe pulse after excitation of the sample with different orientation of the pump polarization, characterised by in-plane angle $\phi$ at a pump energy of \SI{89}{\milli\electronvolt}. The black solid lines are best fits using a sum of two sines, to account for the two frequencies. (d) Amplitude of the Fourier peaks corresponding to the different acoustic pulses for different pump polarization. The dashed lines are the average values and the solid line is a sine fit.}
	\label{Fig:Figure4}
\end{figure}

The Fourier analysis (FFT) of the light-induced coherent strain waves is summarized in Fig.~\ref{Fig:Figure4}a. The FFT amplitudes of the LA and TA strain waves are quantified as a function of the pump photon energy, and summarized in Fig.~\ref{Fig:Figure4}b. We observe that acoustic waves are excited in the optical transparency window, with the amplitude of LA strain being about 6 times larger than the amplitude of the shear wavepacket. Upon reaching the absorption line of the infrared-active phonon we see that the amplitudes of both strain waves start to increase. Surprisingly, in vicinity of the phonon peak, the amplitude of the LA strain wave decreases, while the amplitude of the TA strain peaks, pointing to a competition between the shear deformations and the longitudinal strain. We note that the amplitude of both strain waves increases linearly with pump fluence, both at the resonant and off-resonant pumping as shown in the inset of Fig.~\ref{Fig:Figure4}b. The amplitude of the TA strain remains finite for all the pump polarization and shows a periodic modulation, as seen for the amplitude of the E$_\textrm{g}$ mode (Fig.~\ref{Fig:Figure4}c,d). 

While the generation of a longitudinal strain wave follows straightforwardly from the electrostrictive response of \ce{LaAlO3} and/or optical absorption\cite{artc:Ruello2015, artc:Afanasiev2014}, shear strain must be understood in terms of anisotropic elastic response, both in and out of equilibrium. Indeed, even in equilibrium, (001) \ce{LaAlO3} features an anisotropic surface with respect to the high-symmetry (111) direction that result in off-diagonal elastic constants in the rhombohedral phase. In the supplemental material section II.D we quantify the elastic constants of \ce{LaAlO3} using density functional theory. 

The same calculations show that, out of equilibrium, the displacement along the Raman coordinate driven by the rectification of the phonon field, reinforces the anisotropic elastic response to optical excitation and the generation of shear strain. The relevance of this second mechanism is indicated by the observation that the TA wavefront generation occurs at the expense of the LA wavepacket, indicating that the anisotropic elastic response is further enhanced at the phonon resonance. This observation is also corroborated by the polarization dependence having a component matching the polarization dependence of the E$_\textrm{g}$ mode. 

Our experimental and theoretical analysis uncovers a previously unknown, remarkable feature associated with ultrafast strain engineering. In addition to the displacement along a Raman coordinate and coherent THz atomic vibrations, expected within a nonlinear lattice excitation regime, we observe an efficient generation of shear strain wavepackets. Shear strain following resonant pumping of the crystal lattice in \ce{LaAlO3} is likely to be a key element of the metal-insulator transitions, ultrasonic magnetic dynamics and sonic lattice waves observed in recent years. Tunable shear strain available on the ultrafast timescales can be exploited for material control using a wide array of perovskite anisotropic substrates beyond \ce{LaAlO3}. Since equilibrium shear strain is an important element for ferroelectric\cite{artc:Schlom2014}, flexoelectric\cite{artc:Zubko2013}, piezoelectric and magnetoelectric effects, we envision new opportunities for ultrafast manipulation of collective excitations in solids.

\section{Acknowledgements}
\label{Sec:Acknowledgements}
The authors thank T.C.~van~Thiel for fruitful discussions and G.~Koster for providing a \ce{LaAlO3} sample. This work was partially supported by The Netherlands Organization for Scientific Research (NWO/OCW) as part of the VIDI program, by the EU through the European Research Council, grant No. 677458 (AlterMateria). E.B. and A.S. thank the FRS-FNRS, ARCAIMED project, the C\'{E}CI supercomputer facilities (Grant No. 2.5020.1) and Tier-1 supercomputer of the F\'ed\'eration Wallonie-Bruxelles funded by the Walloon Region (Grant No. 1117545).
\FloatBarrier
\newpage

\clearpage
\setcounter{section}{0}
\setcounter{equation}{0}
\setcounter{figure}{0}
\renewcommand{\thefigure}{S\arabic{figure}}
\renewcommand{\theequation}{S\arabic{equation}}
\renewcommand{\thetable}{S\arabic{table}}
\begin{center}
	\textbf{\large Supplementary Material: \\
		Tunable shear strain from resonantly driven optical phonons}
\end{center}

\section{Temperature dependence of the E$_g$ mode}
\begin{figure}[ht!]
	\centering
	\includegraphics[width = \linewidth]{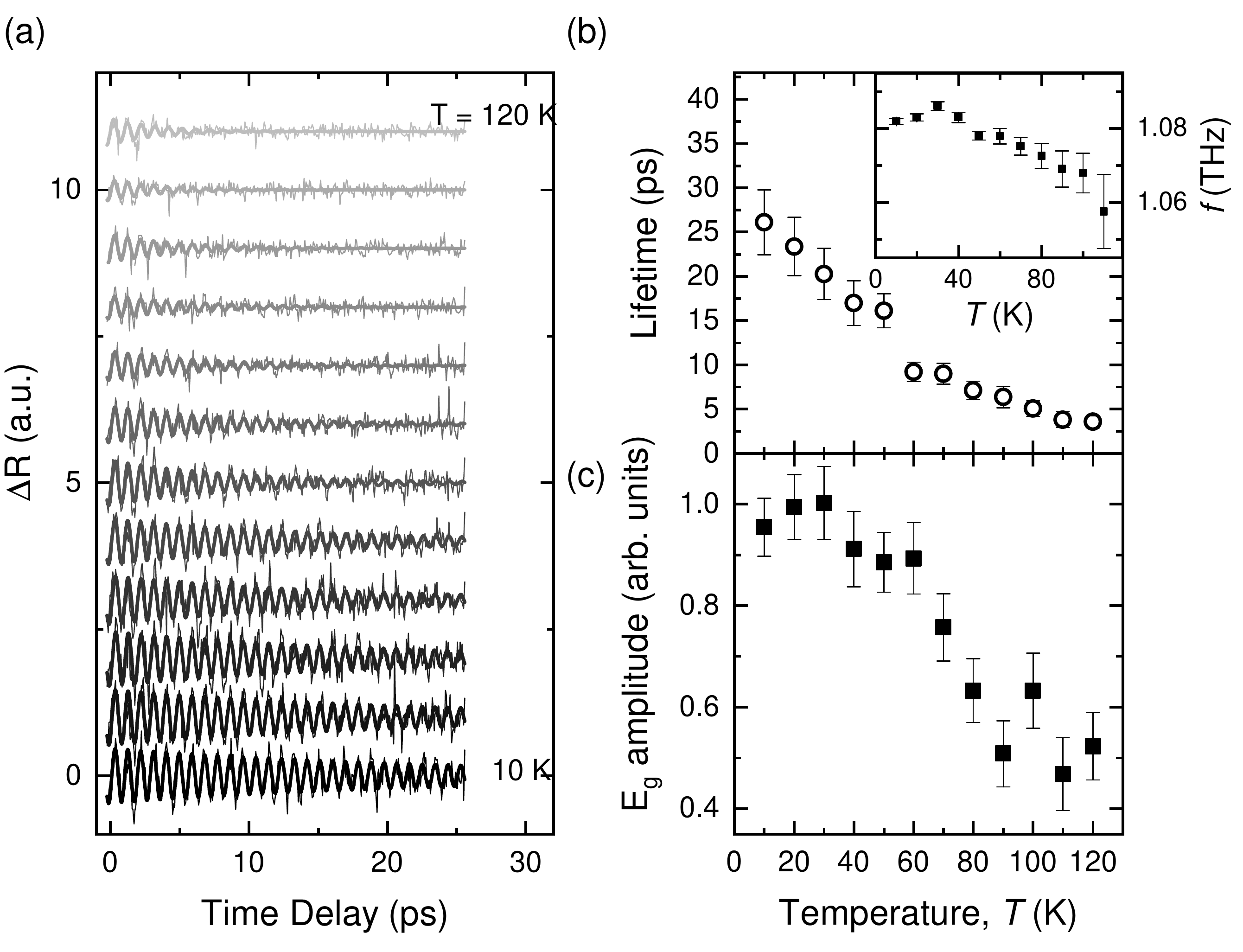}
	\caption{ (a) Time-resolved measurements of the transient change in reflectivity of the probe pulse after excitation of the sample at different temperatures (offset for clarity) with a pump photon energy of \SI{89}{\milli\electronvolt}. The  thicker solid lines represent damped sine fits. (b,c) Lifetime (b), frequency $f$ (inset (b)) and amplitude (c) of the E$_g$ mode oscillations vs. the temperature $T$, extracted using the fits of the data in panel a.}
	\label{Fig:FigureS3}
\end{figure}

In order to confirm that the observed oscillation at \SI{1.1}{\tera\hertz} corresponds to the Raman-active E$_{g}$ mode, we tracked the frequency and lifetime of the excited oscillations as a function of temperature. In Fig.~\ref{Fig:FigureS3} we summarize the experimental findings. Although the frequency of the oscillation $f$ deomnstrates only weak softening upon temeprature increase, its lifetime demonstrates a strong temperature dependence. This behavior as well as exact value of its frequency are cogent hallmarks of the E$_\textrm{g}$ Raman-active vibration being a soft-mode of the rhombohedral-to-pseudocubic structural transition in \ce{LaAlO3} at $T$=527 K.~\cite{artc:Kohmoto2011, artc:Klemens1966}. We also note that in our experimant the amplitude of the E$_g$ mode shows a pronounced decays upon temperature increase.

\section{DFT calculations on nonlinear lattice dynamics}

\subsection{Technical details}
We simulated the $R\bar{3}c$ phase of \ce{LaAlO3} through density functional theory (DFT) \cite{Hohenberg1964,Kohn1965}  as implemented in the ABINIT package (Ver 8.10.2) \cite{Gonze2002,Torrent2008}.
We used Norm-conserving pseudopotentials \cite{Hamann_2013} to account for the interaction of the nuclei and the electrons.
These pseudopotentials were downloaded from the Pseudodojo website \cite{Pspweb}. For \ce{La} we considered $5s$, $5p$, $5d$, $6s$ and $4f$ as valence states and for \ce{Al} and \ce{O} the valence states considered to be $3s$, $3p$ and $2s$, $2p$ respectively.
We used the PBEsol GGA functional for the exchange correlation interaction  \cite{Perdew2008} and all the calculations were done with a 5$\times$5$\times$5 mesh of k-points for sampling of reciprocal space and a cut-off energy on the plane wave expansion of 45 Hartree. 
To calculate the phonons, we used density functional perturbation theory as implemented in ABINIT \cite{DFPT-Gonze,DFPT-Gonzea}.

\subsection{Phonons}
The relaxed lattice parameters are shown in Table \ref{tab:GS-param} together with experimental and DFT calculation values available from literature. There is a good agreement between all of them, besides the usual underestimation of the bandgap given by GGA functionals.

\begin{table}[htbp!]
	\caption{Calculated and experimental ground state parameters of the $R\bar{3}c$ phase of \ce{LaAlO3}. The first column shows the lattice parameter $a$, the second one shows the angle between the lattice vectors, the third one reports the energy band gap and the last column shows the oxygen octahedral rotation angle around the [111] axis.}
	\label{tab:GS-param}
	\begin{tabular}[t]{cccccc@{\hspace*{-\tabcolsep}}}
		\hline
		\hline
		& a (\r{A}) & $\alpha (^{\circ})$ & bandgap (eV) & $\phi_{[111]}(^{\circ})$ \\	
		\hline				
		this work	&	5.337	&	60.2	&	4.16  & 5.97 \\
		DFT	&	5.29\cite{Fredrickson-2016}	&	60.1\cite{Fredrickson-2016}	&	3.88\cite{Fredrickson-2016}  &	5.9\cite{Hatt2010} \\
		EXP	&	5.357\cite{Geller1956}	&	60\cite{Geller1956}	&	5.5\cite{Muller1968} & 6\cite{Geller1956} \\
		\\
		\hline
		\hline	
	\end{tabular}
	
\end{table}

Table \ref{Ph_modes} shows the calculated phonon frequencies for Raman active (E$_g$) and infra-red active (A$_{2u}$ and E$_u$) modes relevant to our study. The calculated phonon frequencies are in good agreement with reported experimental values.

\begin{table}[htbp!]
	\caption{Phonon labels of the $R\bar{3}c$ phase of \ce{LaAlO3} and their frequencies as calculated in this work (column 2), experimentally measured (column 3) and previously DFT calculated (column 4).}
	\label{Ph_modes}
	\begin{tabular}[t]{ccccc@{\hspace*{-\tabcolsep}}}
		\hline
		\hline
		Label & Freq (meV) & Exp	&	DFT \cite{Delugas_2005}	\\	
		\hline				
		
		E$_g$(1)	&	3.82	&	4.22\cite{artc:Scott1969}	&	4.09	\\
		E$_g$(2)	&	18.89	&	18.84\cite{artc:Scott1969}	&	18.10	\\
		E$_g$(3)	&	57.42	&	58.27\cite{artc:Abrashev1999}	&	56.29	\\
		E$_g$(4)	&	59.75	&	60.38\cite{artc:Scott1969}	&	57.90	\\
		&						\\
		
		A$_{2u}$(1)	&	19.75	&		&	20.83	\\
		A$_{2u}$(2)	&	51.04	&		&	50.71	\\
		A$_{2u}$(3)	&	79.78	&		&	77.74	\\
		&						\\
		E$_u$(1)	&	22.74	&	22.56\cite{Calvani_1991}	&	22.19	\\
		E$_u$(2)	&	36.06	&		&	36.82	\\
		E$_u$(3)	&	51.89	&	53.19\cite{Calvani_1991}	&	50.96	\\
		E$_u$(4)	&	60.60	&	62.11\cite{Calvani_1991}	&	59.26	\\
		E$_u$(5)	&	81.66	&	81.45\cite{Calvani_1991}	&	78.98	\\
		\\
		\hline
		\hline	
		
	\end{tabular}
	
\end{table}

\subsection{Phonon-phonon coupling model:}

\begin{figure}[htbp!]
	\includegraphics[width=1\linewidth]{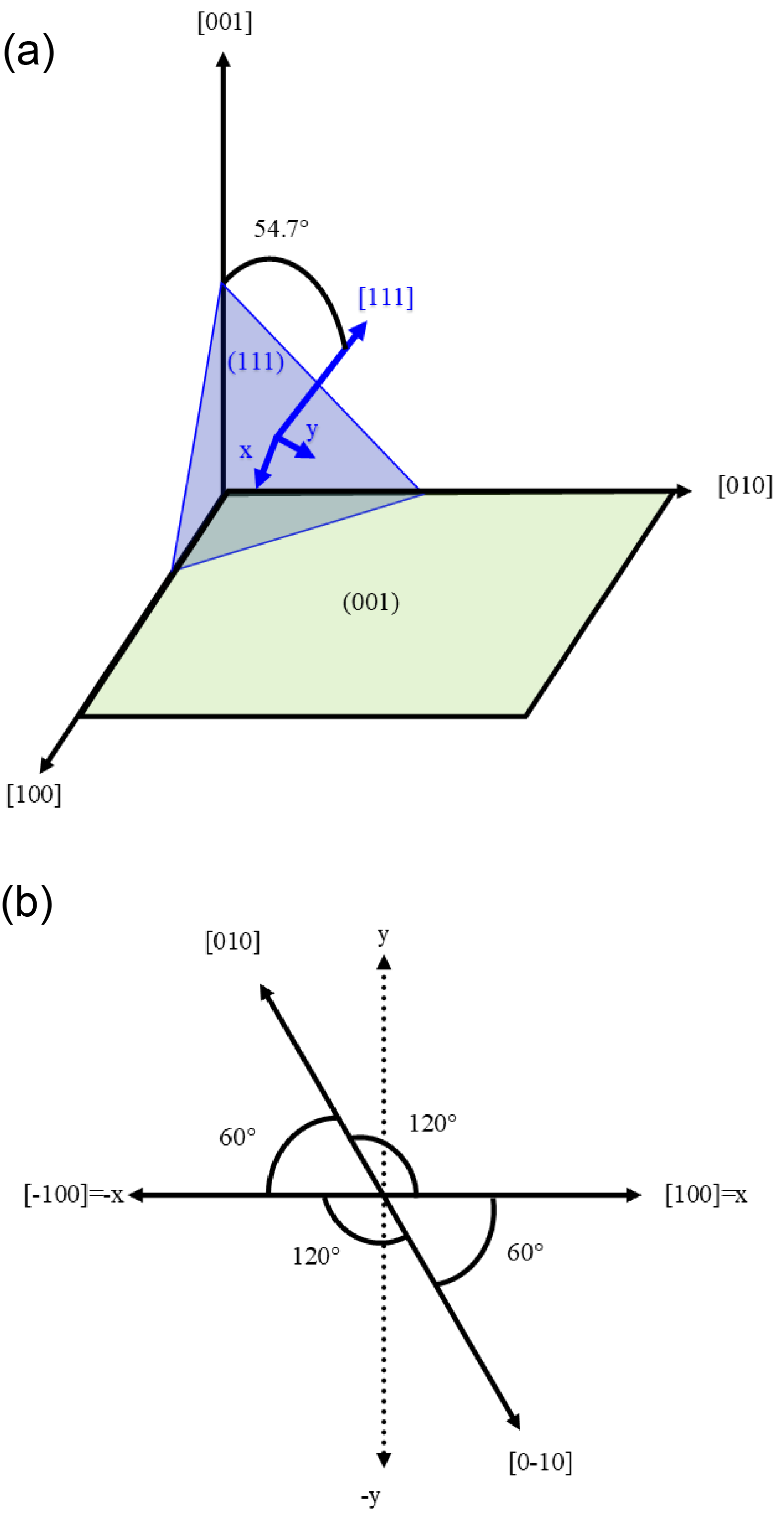} 
	\caption{Schematic view of the lattice planes and directions of pseudocubic and rhombohedral cells. (a) The surface of the sample is [001] oriented but the crystallographic rhombohedral main axis is along the [111] direction. (b) projection of the pseudopseudocubic directions onto rhombohedral coordinates in the (111) plane}
	\label{fig:latt-plnes}
\end{figure}

In our experiment, the resonance amplification of the E$_g$ mode amplitude is observed when the pump photon energy is around \SI{85}{\milli\electronvolt} with the propagation vector in [001] pseudocubic direction. As can be seen from Fig.~\ref{fig:latt-plnes}a this direction makes a $55^{\circ}$ angle with the high symmetry [111] crystallographic axis. This means that the polarization has one component in the [111] direction and one in the (111) plane (from a symmetry point of view, every vector in this space group can be written through the A$_{2u}$-[111] and E$_u$-(111) irreducible representation). Considering the phonon mode energies, the laser can only excite the highest frequency A$_{2u}$ and E$_u$ phonon modes. Hence, we only focus on the excitation of the A$_{2u}$(3) and E$_u$(5) modes (from this point on referred to as A$_{2u}$ and E$_u$ respectively) and their coupling to lower frequency modes. All the other modes are lower in energy and cannot be exited directly by the laser. We further show that although the light can couple to the A$_{2u}$ mode, its excitation cannot account for our experimental findings.

Since the calculations are done in a rhombohedral setting while the orientation of the pump polarization plane is located in the (001) pseudocubic plane, this needs to be taken into account when comparing the results of the calculations to experimental results. Fig~\ref{fig:latt-plnes}b shows the projection of the in-plane vectors of the pseudopseudocubic system to the $x,y$ axes in the (111) plane. Rotating the laser polarization from [100] to [010] direction in pseudocubic setting corresponds to rotating the polarization direction from $x$ in (111) plane by 120$^{\circ}$. Rotating the laser polarization from [010] to [-100] corresponds to rotation of the polarization from 120$^{\circ}$ to -$x$ which is a $60^{\circ}$ angle. This leads to two different functions to describe the connection between results of the experiment to theory due to this difference. 

\subsubsection{Phonon modes}
\begin{figure}
	\includegraphics[width=1\linewidth, height=15cm]{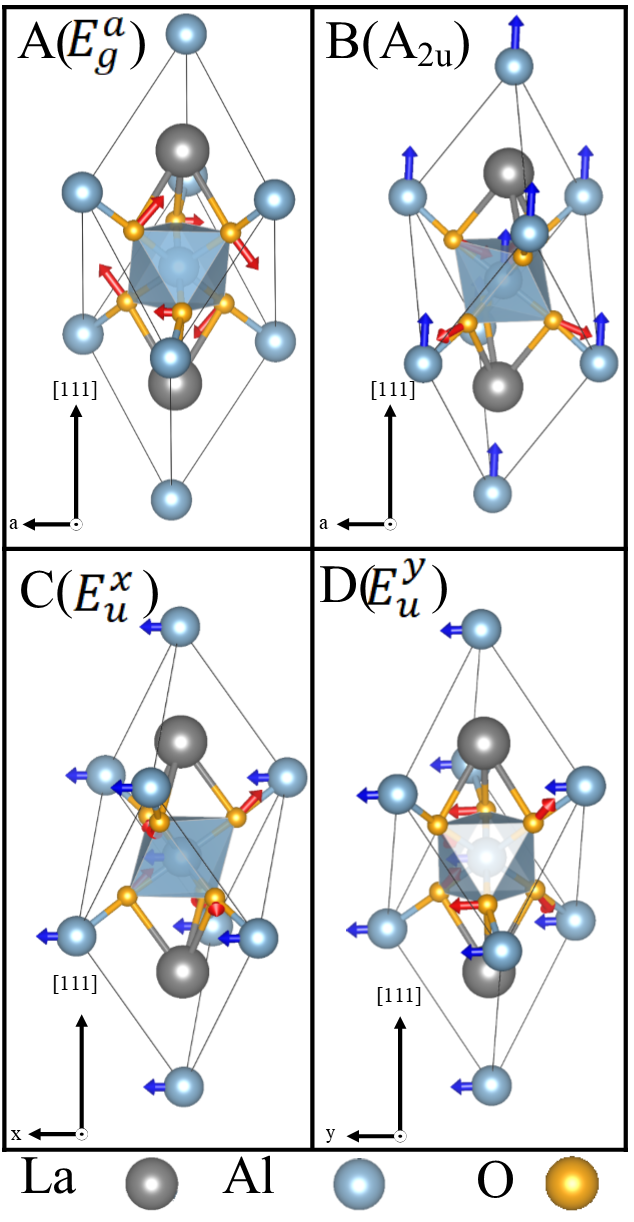} 
	\caption{Schematic pictures of phonon modes vibrations for $E_g$ mode (A), $A_{2u}$ mode (B) and $E_u$ modes (C and D).}
	\label{fig:mod-fig}
\end{figure}

The E$_u$ mode is a two dimensional degenerate phonon mode and depending on the direction on the orientation of the polarization plane of the excitation pulse, we can excite different linear combinations of these two degenerate modes. These modes can be excited within two main axis (orthogonal $x$ and $y$, see Fig.~\ref{fig:latt-plnes}a) or through a linear combination of them.
Hence, for the E$_u$ mode we studied 3 different cases: 
\begin{enumerate}[i]
	\item E$_u^x$ polarized in the $x$ direction in the (111) plane (P1(9)) (Fig.~\ref{fig:mod-fig}c), 
	\item E$_u^y$ polarized in the $y$ direction in the (111) plane (P2(5)) (Fig.~\ref{fig:mod-fig}d),
	\item E$_u^{xy}$ with symmetry C1(1), a linear combination of the E$_u^x$ and E$_u^y$ modes, making a 45 degree angle with $x$ or $y$ directions.
\end{enumerate} 
The A$_{2u}$ mode is not degenerate and is polarized in the [111] direction perpendicular to the (111) plane as presented in Fig~\ref{fig:mod-fig}b.

We studied the coupling of the E$_u$ and A$_{2u}$ mode with several low frequency modes (results not shown here) where we found that all the modes except E$_g$(1) (the lowest frequency one) have negligible coupling to them.
Hence, from this point we focus on the coupling between the E$_u$, A$_{2u}$ and E$_g$(1) modes (from now on referred as E$_g$).

The E$_g$ mode is also doubly degenerate and therefore can have two different symmetries in which the modes could be excited depending on its orientation (E$_g^a$ and E$_g^b$ with P1(15) symmetry shown in Fig~\ref{fig:mod-fig}a and E$_g^{ab}$ with C1(2) symmetry (similar to Fig~\ref{fig:mod-fig}b with different orientation). Here, E$_g^a$ and E$_g^b$ are rotated $60^{\circ}$ with respect to each other.

\subsubsection{E$_u$ and E$_g$ couplings:}
To study the coupling of E$_u$ and E$_g$ modes, we considered 3 different possibilities:

\begin{enumerate}[i]
	\item E$_u^x$ mode with E$_g^a$ and E$_g^{ab}$ modes,
	\item E$_u^y$ mode with E$_g^a$ and E$_g^{ab}$ modes,
	\item E$_u^{xy}$ mode with E$_g^a$ mode.
\end{enumerate} 
For the case of E$_u^x$, E$_u^y$  modes and E$_g^a$ and E$_g^{ab}$ modes, we consider a nonharmonic potential (Eq.~\ref{eq_p1_p1}) and fit the results with DFT calculations. Table \ref{tab:Pot1} shows the fitted coefficient for our simulations.
\begin{equation} \label{eq_p1_p1}
\begin{split} 
V(Q)= a Q_{E_u}^{2}+ b Q_{E_u}^{4}+ a' Q_{E_g}^{2}+b' Q_{E_g}^{3}+\\
c' Q_{E_g}^{4}+\alpha Q_{E_g} Q_{E_u}^2+\beta Q_{E_g}^2 Q_{E_u}^2
\end{split}
\end{equation}

To fit the results of DFT for E$_u^{xy}$ and E$_g^a$, we used Eq.~\ref{eq_p2_p1} and fit the results. For this fit, since we are calculating E$_u^{xy}$ with equal contributions from E$_u^x$ and E$_u^y$ we have the coefficient $a=a"$ and $b=b"$ and the $f(Q_{E_u^1}^{2},Q_{E_u^2}^{2})$ terms can be neglected. The fit parameters for these modes are shown in Table \ref{tab:Pot1}.
\begin{equation} \label{eq_p2_p1}
\begin{split} 
V(Q) & = a Q_{E_u^x}^{2}+ b Q_{E_u^x}^{4}+ a'' Q_{E_u^y}^{2}+ b'' Q_{E_u^y}^{4}+ a' Q_{E_g}^{2}+\\
& b' Q_{E_g}^{3}+ c' Q_{E_g}^{4}+\alpha Q_{E_g} Q_{E_u^x}Q_{E_u^y}+\beta Q_{E_g}^2  Q_{E_u^x}Q_{E_u^y}+\\
& f(Q_{E_u^x}^{2},Q_{E_u^y}^{2})
\end{split}
\end{equation}

\begin{table}[htbp!]
	\caption{Coupling coefficients between E$_u^x$, E$_u^y$, E$_u^{xy}$ modes and E$_g^a$ mode. The units are \si{\electronvolt\per\angstrom\tothe{n}}, where $n$ stands for the order of the coupling.}
	\label{tab:Pot1}
	\begin{tabular}{ccccccccc}
		\hline
		\hline
		&  a	&	b	& $a'$ & $b'$ & $c'$ & $\alpha$  & $\beta$ \\
		\hline
		E$_u^x$ & 13.418 & 11.049 & 0.027 & 0.077 & 1.075  & -1.611  & -3.922 \\
		E$_u^y$ & 13.425 & 10.895 & 0.028 & 0.079 & 1.074  & 1.616  & -4.802  \\
		E$_u^{xy}$ & 6.711 & 5.475 & 0.028 & 0.073 & 1.076  & -0.001  & -4.406  \\
		\hline
	\end{tabular}
	\\
\end{table}

\begin{figure}[htbp!]
	
	\includegraphics[width=0.8\linewidth, height=12cm]{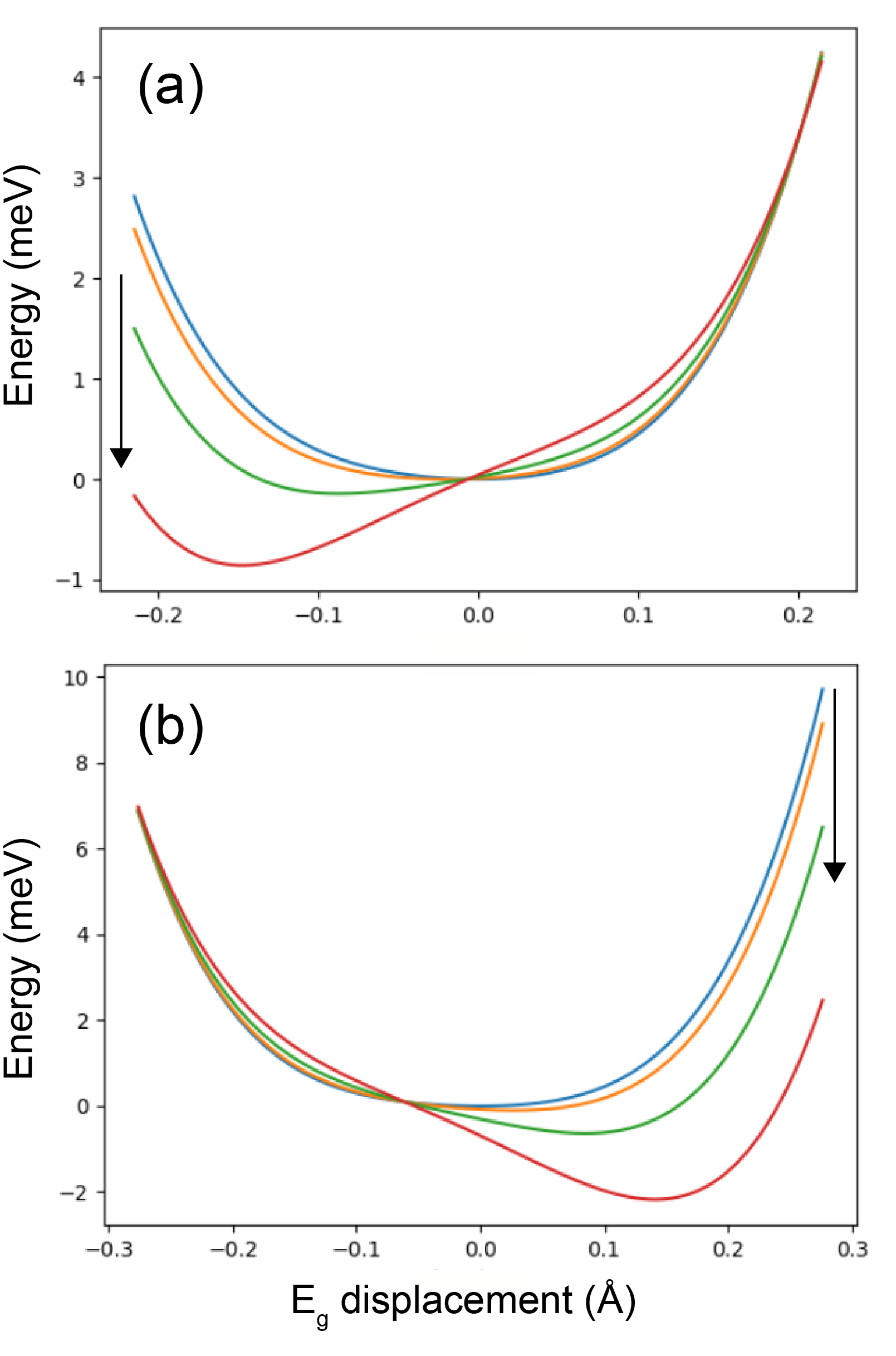} 
	\caption{The potential energy $V$ as a function of the E$_g^a$ mode displacement for different amplitudes of the E$_{u}$ mode (E$_u^x$ (a), E$_u^y$ (b)). The arrows show the direction of increasing E$_{u}$ mode amplitude.}
	\label{fig:mod_28}
\end{figure}

Fig.~\ref{fig:mod_28} shows the evolution of the potential energy $V$ as a function of the E$_g^a$ mode condensation amplitude for different amplitudes of E$_u^x$ (Fig.~\ref{fig:mod_28}a) and E$_u^y$ (Fig.~\ref{fig:mod_28}b) modes.
The modes E$_u^x$ and E$_u^y$ tend to displace the minimum of the E$_g^a$ mode toward different directions. These behaviours are due to opposite signs of the $\alpha$ coefficients that couples the two E$_u^x$ and E$_u^y$ modes to the E$_g^a$ mode. Hence, this shifts the minimum of the energy in two different directions, which could be the reason why we have oscillation of the E$_g$ modes in the experiments that has different phase for two modes in $0^{\circ}$ (exciting E$_u^x$ mode) and $90^{\circ}$ polarization (mainly exciting the E$_u^y$ mode) of the laser. A minimum different from 0, means that excitation of the E$_u^x$ and E$_u^y$ modes quasi-statically freezes the E$_g^a$ mode (lowering the symmetry), which is the characteristic of nonlinear phononics.\cite{artc:Forst2011_2}

For the E$_u^{xy}$ mode (which is a linear combination of the E$_u^x$ and E$_u^y$ mode with equal contribution from each one) the $\alpha$ parameter is zero and the higher order coupling are not large enough to create any considerable dynamics. This situation occurs when the polarization component of the laser in the (111) plane is exactly between the $x$ and $y$ axis.
To further study the effects of polarization direction of the laser on E$_g^a$ mode, we have studied the evolution of the $\alpha$ parameter with respect to different polarization angles in the (001) plane. Figure 3c in the main text shows the results with $\alpha$ changing in an oscillatory manner. The reason that the function consists of two parts, is because the polarization in the (001) plane has to be projected on the (111) plane, see Fig.~\ref{fig:latt-plnes}b. 

The coupling of different E$_u$ modes with  E$_g^b$ mode is also studied. The results of these studies are the same as the one presented for  E$_g^a$ case with the difference that the $b'$ coefficient is positive in the E$_g^a$ case while it is negative in the E$_g^b$ case. The associated effects of E$_u^x$ or E$_u^y$ modes coming from this coefficient is small compared to the one given  by the $\alpha$ coefficient. 

\subsubsection{Fluence dependence of the E$_g$ mode}
\begin{figure}[ht!]
	\centering
	\includegraphics[width = \linewidth]{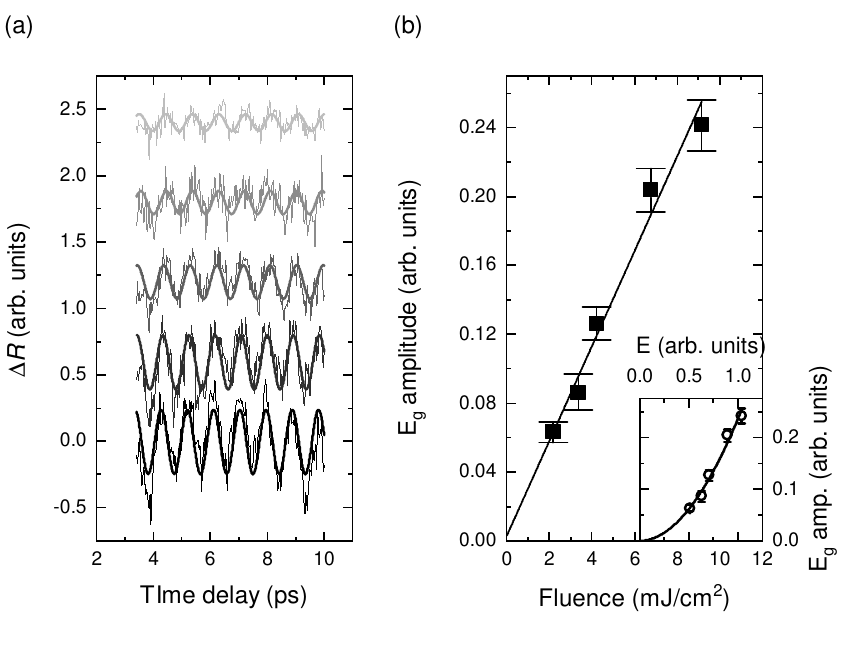}
	\caption{ (a) Time-resolved measurements of the transient change in reflectivity of the probe pulse after excitation of the sample at a pump energy of \SI{89}{\milli\electronvolt} with different fluences (offset for clarity). The  thicker solid lines represent sine fits. (b) Amplitude of the E$_g$ mode oscillations vs. pump fluence, extracted using the fits of the data in panel a. Inset: Amplitude of the E$_g$ mode vs. pump electric field $E$ with a power fit giving $E^{1.93 \pm 0.15}$.}
	\label{Fig:FigureS1}
\end{figure}

We measured the dependence of the observed oscillations corresponding to the non-polar E$_g$ mode as a function of the pump fluence at the resonance conditions (with the pump photon energy of 83 meV). As shown in Fig.~\ref{Fig:FigureS1}, the oscillation amplitude depends linearly on the pump fluence. This observation, combined with the resonant character, indicates a quadratic dependence of the E$_g$ mode amplitude on amplitude of the pump driven E$_u$ mode.

\subsubsection{A$_{2u}$ and E$_g$ coupling:}

The A$_{2u}$ mode is single dimensional and it could couple to different directions of the E$_g$ mode (E$_g^a$ and E$_g^b$ and E$_g^{ab}$ modes).
We fitted Eq. \ref{eq_p1_p1} to study the A$_{2u}$ mode coupling where $\alpha$ is equal to zero (it is not allowed by symmetry).
The resulting calculated coefficients are shown in Tab \ref{tab:Pot2}.

\begin{table}[htbp!]
	\caption{coupling coefficients between the A$_{2u}$ mode and E$_g^a$, E$_g^b$ and E$_g^{ab}$ modes. The units are \si{\electronvolt\per\angstrom\tothe{n}}, where $n$ stands for the order of the coupling.}
	\label{tab:Pot2}
	\begin{tabular}{cccccccc}
		\hline
		\hline
		&  a	&	b	& $a'$ & $b'$ & $c'$ &  $\beta$ \\
		\hline
		E$_g^a$ & 12.81 & 8.01 & 0.027 & 0.078 & 1.059    & -3.161 \\
		E$_g^b$ & 12.81 & 8.00 & 0.027 & -0.084 & 1.062    & -3.159  \\
		E$_g^{ab}$ & 12.815 & 8.020 & 0.027 & -0.012 & -0.012    & -3.167  \\
		\hline
	\end{tabular}
	\\
\end{table}

In this case $\beta$ is the coefficient that is governing the change in energy surface of the E$_g$ mode but it is very small with respect to the E$_u$ case such that the dynamics resulting from the A$_{2u}$ excitation can be neglected.

\subsection{Strain waves:}
In the experiment, both longitudinal acoustic (LA) and transverse acoustic (TA) waves are observed after resonant optical phonon excitation. 
Although exciting the LA mode is straightforward, exciting the TA mode is not evident  in isotropic materials.
We have studied two possible origins to excite the TA mode: (i) by having the crystal with off-axis orientation with respect to the sample surface or (ii) through the phonon-phonon coupling that induces anisotropic strain. 

In case (i), the experimental set-up is such that the LaAlO$_3$ sample surface is oriented in the pseudocubic [001] direction, which makes an angle of $55^{\circ}$ with respect to the proper rhombohedral [111] high symmetry direction (as can be seen in Fig.~\ref{fig:latt-plnes}a).
This is evidenced from the calculated elastic constants of both the pseudocubic and the rhombohedral phases reported in Table \ref{tab:el-const}. We can see that the transition from pseudocubic to $R\bar{3}c$ phase creates anisotropy in the pseudocubic elasticity with a pronounced splitting between the rhombohedral $xy$ directions and the $z$ direction. In this case the shear strain will be exited due to anisotropy arising from off-axis orientation of the crystal.

In the second case (ii) the phonon-phonon coupling induced by the laser, excites the E$_g$ mode, which in turn can induce an out-of-equilibrium anisotropy into the elasticity. To further show the effect of this dynamically induced phonon induced stress, we relaxed the lattice parameters by keeping the E$_g$ mode frozen in the crystal and compared the resulting elastic constants with respect to the ones of pseudocubic and rhombohedral LaAlO$_3$ (Table \ref{tab:el-const}). 
It is seen that the anisotropy arising from the E$_g$ mode condensation has a strong impact on the C$_\textrm{35}$ and C$_\textrm{51}$ elastic constants. These off-diagonal constants indeed couple longitudinal and shear strain likely contributing to the strain conversion when energy of the pump photon is swiped across the E$_\textrm{u}$ phonon resonance.

\begin{table}[htbp!]
	\caption{Elastic constants of LaAlO$_3$ (in Cartesian directions) for the pseudocubic, rhombohedral ($R\bar{3}c$) and rhombohedral with the E$_g$ mode condensed. The amplitude of the distortion is \SI{0.13}{\angstrom}. Units are in (GPa)}
	\label{tab:el-const}
	\begin{tabular}[t]{cccc@{\hspace*{-\tabcolsep}}}
		\hline
		\hline
		& pseudocubic & Rhombo & Rhombo+E$_g$  \\	
		\hline				
		C$_{11}$	 &	 366  & 404  &  410 \\
		C$_{22}$	 &	 366  & 404  &  394 \\
		C$_{33}$ &	 366  & 406  &  406 \\
		C$_{12}$ & 130   &  124   &  124 \\
		C$_{23}$ & 130   &  98   &  97 \\
		C$_{31}$ & 130   &  98   &  100 \\
		C$_{44}$ &	 0  & 124  &  123 \\
		C$_{55}$ &	 0  & 124  &  124 \\
		C$_{66}$ &	 0  & 140  &  140 \\
		C$_{46}$ & 0   &  -22   &  -23 \\
		C$_{51}$ & 0   &  22   &  18 \\
		C$_{52}$ & 0   &  -22   &  -22 \\
		C$_{35}$ & 0   &  0   &  -4 \\
		\hline
		\hline	
	\end{tabular}
\end{table}

\FloatBarrier


\bibliography{biblio}

\begin{thebibliography}{42}%
\makeatletter
\providecommand \@ifxundefined [1]{%
 \@ifx{#1\undefined}
}%
\providecommand \@ifnum [1]{%
 \ifnum #1\expandafter \@firstoftwo
 \else \expandafter \@secondoftwo
 \fi
}%
\providecommand \@ifx [1]{%
 \ifx #1\expandafter \@firstoftwo
 \else \expandafter \@secondoftwo
 \fi
}%
\providecommand \natexlab [1]{#1}%
\providecommand \enquote  [1]{``#1''}%
\providecommand \bibnamefont  [1]{#1}%
\providecommand \bibfnamefont [1]{#1}%
\providecommand \citenamefont [1]{#1}%
\providecommand \href@noop [0]{\@secondoftwo}%
\providecommand \href [0]{\begingroup \@sanitize@url \@href}%
\providecommand \@href[1]{\@@startlink{#1}\@@href}%
\providecommand \@@href[1]{\endgroup#1\@@endlink}%
\providecommand \@sanitize@url [0]{\catcode `\\12\catcode `\$12\catcode
  `\&12\catcode `\#12\catcode `\^12\catcode `\_12\catcode `\%12\relax}%
\providecommand \@@startlink[1]{}%
\providecommand \@@endlink[0]{}%
\providecommand \url  [0]{\begingroup\@sanitize@url \@url }%
\providecommand \@url [1]{\endgroup\@href {#1}{\urlprefix }}%
\providecommand \urlprefix  [0]{URL }%
\providecommand \Eprint [0]{\href }%
\providecommand \doibase [0]{http://dx.doi.org/}%
\providecommand \selectlanguage [0]{\@gobble}%
\providecommand \bibinfo  [0]{\@secondoftwo}%
\providecommand \bibfield  [0]{\@secondoftwo}%
\providecommand \translation [1]{[#1]}%
\providecommand \BibitemOpen [0]{}%
\providecommand \bibitemStop [0]{}%
\providecommand \bibitemNoStop [0]{.\EOS\space}%
\providecommand \EOS [0]{\spacefactor3000\relax}%
\providecommand \BibitemShut  [1]{\csname bibitem#1\endcsname}%
\let\auto@bib@innerbib\@empty
\bibitem [{\citenamefont {Schlom}\ \emph {et~al.}(2014)\citenamefont {Schlom},
  \citenamefont {Chen}, \citenamefont {Fennie}, \citenamefont {Gopalan},
  \citenamefont {Muller}, \citenamefont {Pan}, \citenamefont {Ramesh},\ and\
  \citenamefont {Uecker}}]{artc:Schlom2014}%
  \BibitemOpen
  \bibfield  {author} {\bibinfo {author} {\bibfnamefont {D.~G.}\ \bibnamefont
  {Schlom}}, \bibinfo {author} {\bibfnamefont {L.~Q.}\ \bibnamefont {Chen}},
  \bibinfo {author} {\bibfnamefont {C.~J.}\ \bibnamefont {Fennie}}, \bibinfo
  {author} {\bibfnamefont {V.}~\bibnamefont {Gopalan}}, \bibinfo {author}
  {\bibfnamefont {D.~A.}\ \bibnamefont {Muller}}, \bibinfo {author}
  {\bibfnamefont {X.}~\bibnamefont {Pan}}, \bibinfo {author} {\bibfnamefont
  {R.}~\bibnamefont {Ramesh}}, \ and\ \bibinfo {author} {\bibfnamefont
  {R.}~\bibnamefont {Uecker}},\ }\bibfield  {title} {\enquote {\bibinfo {title}
  {Elastic strain engineering of ferroic oxides},}\ }\href {\doibase
  10.1557/mrs.2014.1} {\bibfield  {journal} {\bibinfo  {journal} {{MRS}
  {B}ulletin}\ }\textbf {\bibinfo {volume} {39}},\ \bibinfo {pages} {118--130}
  (\bibinfo {year} {2014})}\BibitemShut {NoStop}%
\bibitem [{\citenamefont {Mundy}\ \emph {et~al.}(2016)\citenamefont {Mundy},
  \citenamefont {Brooks}, \citenamefont {Holtz}, \citenamefont {Moyer},
  \citenamefont {Das}, \citenamefont {R{\'{e}}bola}, \citenamefont {Heron},
  \citenamefont {Clarkson}, \citenamefont {Disseler}, \citenamefont {Liu},
  \citenamefont {Farhan}, \citenamefont {Held}, \citenamefont {Hovden},
  \citenamefont {Padgett}, \citenamefont {Mao}, \citenamefont {Paik},
  \citenamefont {Misra}, \citenamefont {Kourkoutis}, \citenamefont {Arenholz},
  \citenamefont {Scholl}, \citenamefont {Borchers}, \citenamefont {Ratcliff},
  \citenamefont {Ramesh}, \citenamefont {Fennie}, \citenamefont {Schiffer},
  \citenamefont {Muller},\ and\ \citenamefont {Schlom}}]{Mundy2016}%
  \BibitemOpen
  \bibfield  {author} {\bibinfo {author} {\bibfnamefont {J.~A.}\ \bibnamefont
  {Mundy}}, \bibinfo {author} {\bibfnamefont {C.~M.}\ \bibnamefont {Brooks}},
  \bibinfo {author} {\bibfnamefont {M.~E.}\ \bibnamefont {Holtz}}, \bibinfo
  {author} {\bibfnamefont {J.~A.}\ \bibnamefont {Moyer}}, \bibinfo {author}
  {\bibfnamefont {H.}~\bibnamefont {Das}}, \bibinfo {author} {\bibfnamefont
  {A.~F.}\ \bibnamefont {R{\'{e}}bola}}, \bibinfo {author} {\bibfnamefont
  {J.~T.}\ \bibnamefont {Heron}}, \bibinfo {author} {\bibfnamefont {J.~D.}\
  \bibnamefont {Clarkson}}, \bibinfo {author} {\bibfnamefont {S.~M.}\
  \bibnamefont {Disseler}}, \bibinfo {author} {\bibfnamefont {Z.}~\bibnamefont
  {Liu}}, \bibinfo {author} {\bibfnamefont {A.}~\bibnamefont {Farhan}},
  \bibinfo {author} {\bibfnamefont {R.}~\bibnamefont {Held}}, \bibinfo {author}
  {\bibfnamefont {R.}~\bibnamefont {Hovden}}, \bibinfo {author} {\bibfnamefont
  {E.}~\bibnamefont {Padgett}}, \bibinfo {author} {\bibfnamefont
  {Q.}~\bibnamefont {Mao}}, \bibinfo {author} {\bibfnamefont {H.}~\bibnamefont
  {Paik}}, \bibinfo {author} {\bibfnamefont {R.}~\bibnamefont {Misra}},
  \bibinfo {author} {\bibfnamefont {L.~F.}\ \bibnamefont {Kourkoutis}},
  \bibinfo {author} {\bibfnamefont {E.}~\bibnamefont {Arenholz}}, \bibinfo
  {author} {\bibfnamefont {A.}~\bibnamefont {Scholl}}, \bibinfo {author}
  {\bibfnamefont {J.~A.}\ \bibnamefont {Borchers}}, \bibinfo {author}
  {\bibfnamefont {W.~D.}\ \bibnamefont {Ratcliff}}, \bibinfo {author}
  {\bibfnamefont {R.}~\bibnamefont {Ramesh}}, \bibinfo {author} {\bibfnamefont
  {C.~J.}\ \bibnamefont {Fennie}}, \bibinfo {author} {\bibfnamefont
  {P.}~\bibnamefont {Schiffer}}, \bibinfo {author} {\bibfnamefont {D.~A.}\
  \bibnamefont {Muller}}, \ and\ \bibinfo {author} {\bibfnamefont {D.~G.}\
  \bibnamefont {Schlom}},\ }\bibfield  {title} {\enquote {\bibinfo {title}
  {Atomically engineered ferroic layers yield a room-temperature
  magnetoelectric multiferroic},}\ }\href {\doibase 10.1038/nature19343}
  {\bibfield  {journal} {\bibinfo  {journal} {Nature}\ }\textbf {\bibinfo
  {volume} {537}},\ \bibinfo {pages} {523--527} (\bibinfo {year}
  {2016})}\BibitemShut {NoStop}%
\bibitem [{\citenamefont {Haeni}\ \emph {et~al.}(2004)\citenamefont {Haeni},
  \citenamefont {Irvin}, \citenamefont {Chang}, \citenamefont {Uecker},
  \citenamefont {Reiche}, \citenamefont {Li}, \citenamefont {Choudhury},
  \citenamefont {Tian}, \citenamefont {Hawley}, \citenamefont {Craigo} \emph
  {et~al.}}]{artc:Haeni2004}%
  \BibitemOpen
  \bibfield  {author} {\bibinfo {author} {\bibfnamefont {J.}~\bibnamefont
  {Haeni}}, \bibinfo {author} {\bibfnamefont {P.}~\bibnamefont {Irvin}},
  \bibinfo {author} {\bibfnamefont {W.}~\bibnamefont {Chang}}, \bibinfo
  {author} {\bibfnamefont {R.}~\bibnamefont {Uecker}}, \bibinfo {author}
  {\bibfnamefont {P.}~\bibnamefont {Reiche}}, \bibinfo {author} {\bibfnamefont
  {Y.}~\bibnamefont {Li}}, \bibinfo {author} {\bibfnamefont {S.}~\bibnamefont
  {Choudhury}}, \bibinfo {author} {\bibfnamefont {W.}~\bibnamefont {Tian}},
  \bibinfo {author} {\bibfnamefont {M.}~\bibnamefont {Hawley}}, \bibinfo
  {author} {\bibfnamefont {B.}~\bibnamefont {Craigo}},  \emph {et~al.},\
  }\bibfield  {title} {\enquote {\bibinfo {title} {Room-temperature
  ferroelectricity in strained {SrTiO$_3$}},}\ }\href@noop {} {\bibfield
  {journal} {\bibinfo  {journal} {Nature}\ }\textbf {\bibinfo {volume} {430}},\
  \bibinfo {pages} {758--761} (\bibinfo {year} {2004})}\BibitemShut {NoStop}%
\bibitem [{\citenamefont {Schlom}\ \emph {et~al.}(2007)\citenamefont {Schlom},
  \citenamefont {Chen}, \citenamefont {Eom}, \citenamefont {Rabe},
  \citenamefont {Streiffer},\ and\ \citenamefont {Triscone}}]{artc:Schlom2007}%
  \BibitemOpen
  \bibfield  {author} {\bibinfo {author} {\bibfnamefont {D.~G.}\ \bibnamefont
  {Schlom}}, \bibinfo {author} {\bibfnamefont {L.-Q.}\ \bibnamefont {Chen}},
  \bibinfo {author} {\bibfnamefont {C.-B.}\ \bibnamefont {Eom}}, \bibinfo
  {author} {\bibfnamefont {K.~M.}\ \bibnamefont {Rabe}}, \bibinfo {author}
  {\bibfnamefont {S.~K.}\ \bibnamefont {Streiffer}}, \ and\ \bibinfo {author}
  {\bibfnamefont {J.-M.}\ \bibnamefont {Triscone}},\ }\bibfield  {title}
  {\enquote {\bibinfo {title} {Strain tuning of ferroelectric thin films},}\
  }\href {\doibase 10.1146/annurev.matsci.37.061206.113016} {\bibfield
  {journal} {\bibinfo  {journal} {Annual Review of Materials Research}\
  }\textbf {\bibinfo {volume} {37}},\ \bibinfo {pages} {589--626} (\bibinfo
  {year} {2007})}\BibitemShut {NoStop}%
\bibitem [{\citenamefont {Sando}\ \emph {et~al.}(2013)\citenamefont {Sando},
  \citenamefont {Agbelele}, \citenamefont {Rahmedov}, \citenamefont {Liu},
  \citenamefont {Rovillain}, \citenamefont {Toulouse}, \citenamefont {Infante},
  \citenamefont {Pyatakov}, \citenamefont {Fusil}, \citenamefont {Jacquet}
  \emph {et~al.}}]{artc:Sando2013}%
  \BibitemOpen
  \bibfield  {author} {\bibinfo {author} {\bibfnamefont {D.}~\bibnamefont
  {Sando}}, \bibinfo {author} {\bibfnamefont {A.}~\bibnamefont {Agbelele}},
  \bibinfo {author} {\bibfnamefont {D.}~\bibnamefont {Rahmedov}}, \bibinfo
  {author} {\bibfnamefont {J.}~\bibnamefont {Liu}}, \bibinfo {author}
  {\bibfnamefont {P.}~\bibnamefont {Rovillain}}, \bibinfo {author}
  {\bibfnamefont {C.}~\bibnamefont {Toulouse}}, \bibinfo {author}
  {\bibfnamefont {I.}~\bibnamefont {Infante}}, \bibinfo {author} {\bibfnamefont
  {A.}~\bibnamefont {Pyatakov}}, \bibinfo {author} {\bibfnamefont
  {S.}~\bibnamefont {Fusil}}, \bibinfo {author} {\bibfnamefont
  {E.}~\bibnamefont {Jacquet}},  \emph {et~al.},\ }\bibfield  {title} {\enquote
  {\bibinfo {title} {Crafting the magnonic and spintronic response of
  {BiFeO$_3$} films by epitaxial strain},}\ }\href@noop {} {\bibfield
  {journal} {\bibinfo  {journal} {Nature materials}\ }\textbf {\bibinfo
  {volume} {12}},\ \bibinfo {pages} {641--646} (\bibinfo {year}
  {2013})}\BibitemShut {NoStop}%
\bibitem [{\citenamefont {Caviglia}\ \emph {et~al.}(2012)\citenamefont
  {Caviglia}, \citenamefont {Scherwitzl}, \citenamefont {Popovich},
  \citenamefont {Hu}, \citenamefont {Bromberger}, \citenamefont {Singla},
  \citenamefont {Mitrano}, \citenamefont {Hoffmann}, \citenamefont {Kaiser},
  \citenamefont {Zubko}, \citenamefont {Gariglio}, \citenamefont {Triscone},
  \citenamefont {F{\"{o}}rst},\ and\ \citenamefont
  {Cavalleri}}]{artc:Caviglia2012}%
  \BibitemOpen
  \bibfield  {author} {\bibinfo {author} {\bibfnamefont {A.~D.}\ \bibnamefont
  {Caviglia}}, \bibinfo {author} {\bibfnamefont {R.}~\bibnamefont
  {Scherwitzl}}, \bibinfo {author} {\bibfnamefont {P.}~\bibnamefont
  {Popovich}}, \bibinfo {author} {\bibfnamefont {W.}~\bibnamefont {Hu}},
  \bibinfo {author} {\bibfnamefont {H.}~\bibnamefont {Bromberger}}, \bibinfo
  {author} {\bibfnamefont {R.}~\bibnamefont {Singla}}, \bibinfo {author}
  {\bibfnamefont {M.}~\bibnamefont {Mitrano}}, \bibinfo {author} {\bibfnamefont
  {M.~C.}\ \bibnamefont {Hoffmann}}, \bibinfo {author} {\bibfnamefont
  {S.}~\bibnamefont {Kaiser}}, \bibinfo {author} {\bibfnamefont
  {P.}~\bibnamefont {Zubko}}, \bibinfo {author} {\bibfnamefont
  {S.}~\bibnamefont {Gariglio}}, \bibinfo {author} {\bibfnamefont {J.~M.}\
  \bibnamefont {Triscone}}, \bibinfo {author} {\bibfnamefont {M.}~\bibnamefont
  {F{\"{o}}rst}}, \ and\ \bibinfo {author} {\bibfnamefont {A.}~\bibnamefont
  {Cavalleri}},\ }\bibfield  {title} {\enquote {\bibinfo {title} {Ultrafast
  strain engineering in complex oxide heterostructures},}\ }\href {\doibase
  10.1103/PhysRevLett.108.136801} {\bibfield  {journal} {\bibinfo  {journal}
  {Phys. Rev. Lett.}\ }\textbf {\bibinfo {volume} {108}} (\bibinfo {year}
  {2012}),\ 10.1103/PhysRevLett.108.136801}\BibitemShut {NoStop}%
\bibitem [{\citenamefont {F{\"{o}}rst}\ \emph {et~al.}(2015)\citenamefont
  {F{\"{o}}rst}, \citenamefont {Caviglia}, \citenamefont {Scherwitzl},
  \citenamefont {Mankowsky}, \citenamefont {Zubko}, \citenamefont {Khanna},
  \citenamefont {Bromberger}, \citenamefont {Wilkins}, \citenamefont {Chuang},
  \citenamefont {Lee}, \citenamefont {Schlotter}, \citenamefont {Turner},
  \citenamefont {Dakovski}, \citenamefont {Minitti}, \citenamefont {Robinson},
  \citenamefont {Clark}, \citenamefont {Jaksch}, \citenamefont {Triscone},
  \citenamefont {Hill}, \citenamefont {Dhesi},\ and\ \citenamefont
  {Cavalleri}}]{artc:Forst2015}%
  \BibitemOpen
  \bibfield  {author} {\bibinfo {author} {\bibfnamefont {M.}~\bibnamefont
  {F{\"{o}}rst}}, \bibinfo {author} {\bibfnamefont {A.~D.}\ \bibnamefont
  {Caviglia}}, \bibinfo {author} {\bibfnamefont {R.}~\bibnamefont
  {Scherwitzl}}, \bibinfo {author} {\bibfnamefont {R.}~\bibnamefont
  {Mankowsky}}, \bibinfo {author} {\bibfnamefont {P.}~\bibnamefont {Zubko}},
  \bibinfo {author} {\bibfnamefont {V.}~\bibnamefont {Khanna}}, \bibinfo
  {author} {\bibfnamefont {H.}~\bibnamefont {Bromberger}}, \bibinfo {author}
  {\bibfnamefont {S.~B.}\ \bibnamefont {Wilkins}}, \bibinfo {author}
  {\bibfnamefont {Y.~D.}\ \bibnamefont {Chuang}}, \bibinfo {author}
  {\bibfnamefont {W.~S.}\ \bibnamefont {Lee}}, \bibinfo {author} {\bibfnamefont
  {W.~F.}\ \bibnamefont {Schlotter}}, \bibinfo {author} {\bibfnamefont {J.~J.}\
  \bibnamefont {Turner}}, \bibinfo {author} {\bibfnamefont {G.~L.}\
  \bibnamefont {Dakovski}}, \bibinfo {author} {\bibfnamefont {M.~P.}\
  \bibnamefont {Minitti}}, \bibinfo {author} {\bibfnamefont {J.}~\bibnamefont
  {Robinson}}, \bibinfo {author} {\bibfnamefont {S.~R.}\ \bibnamefont {Clark}},
  \bibinfo {author} {\bibfnamefont {D.}~\bibnamefont {Jaksch}}, \bibinfo
  {author} {\bibfnamefont {J.~M.}\ \bibnamefont {Triscone}}, \bibinfo {author}
  {\bibfnamefont {J.~P.}\ \bibnamefont {Hill}}, \bibinfo {author}
  {\bibfnamefont {S.~S.}\ \bibnamefont {Dhesi}}, \ and\ \bibinfo {author}
  {\bibfnamefont {A.}~\bibnamefont {Cavalleri}},\ }\bibfield  {title} {\enquote
  {\bibinfo {title} {Spatially resolved ultrafast magnetic dynamics initiated
  at a complex oxide heterointerface},}\ }\href {\doibase 10.1038/nmat4341}
  {\bibfield  {journal} {\bibinfo  {journal} {Nature Materials}\ }\textbf
  {\bibinfo {volume} {14}},\ \bibinfo {pages} {883--888} (\bibinfo {year}
  {2015})},\ \Eprint {http://arxiv.org/abs/1505.00601} {1505.00601}
  \BibitemShut {NoStop}%
\bibitem [{\citenamefont {F{\"{o}}rst}\ \emph {et~al.}(2017)\citenamefont
  {F{\"{o}}rst}, \citenamefont {Beyerlein}, \citenamefont {Mankowsky},
  \citenamefont {Hu}, \citenamefont {Mattoni}, \citenamefont {Catalano},
  \citenamefont {Gibert}, \citenamefont {Yefanov}, \citenamefont {Clark},
  \citenamefont {Frano}, \citenamefont {Glownia}, \citenamefont {Chollet},
  \citenamefont {Lemke}, \citenamefont {Moser}, \citenamefont {Collins},
  \citenamefont {Dhesi}, \citenamefont {Caviglia}, \citenamefont {Triscone},\
  and\ \citenamefont {Cavalleri}}]{artc:Forst2017}%
  \BibitemOpen
  \bibfield  {author} {\bibinfo {author} {\bibfnamefont {M.}~\bibnamefont
  {F{\"{o}}rst}}, \bibinfo {author} {\bibfnamefont {K.~R.}\ \bibnamefont
  {Beyerlein}}, \bibinfo {author} {\bibfnamefont {R.}~\bibnamefont
  {Mankowsky}}, \bibinfo {author} {\bibfnamefont {W.}~\bibnamefont {Hu}},
  \bibinfo {author} {\bibfnamefont {G.}~\bibnamefont {Mattoni}}, \bibinfo
  {author} {\bibfnamefont {S.}~\bibnamefont {Catalano}}, \bibinfo {author}
  {\bibfnamefont {M.}~\bibnamefont {Gibert}}, \bibinfo {author} {\bibfnamefont
  {O.}~\bibnamefont {Yefanov}}, \bibinfo {author} {\bibfnamefont {J.~N.}\
  \bibnamefont {Clark}}, \bibinfo {author} {\bibfnamefont {A.}~\bibnamefont
  {Frano}}, \bibinfo {author} {\bibfnamefont {J.~M.}\ \bibnamefont {Glownia}},
  \bibinfo {author} {\bibfnamefont {M.}~\bibnamefont {Chollet}}, \bibinfo
  {author} {\bibfnamefont {H.}~\bibnamefont {Lemke}}, \bibinfo {author}
  {\bibfnamefont {B.}~\bibnamefont {Moser}}, \bibinfo {author} {\bibfnamefont
  {S.~P.}\ \bibnamefont {Collins}}, \bibinfo {author} {\bibfnamefont {S.~S.}\
  \bibnamefont {Dhesi}}, \bibinfo {author} {\bibfnamefont {A.~D.}\ \bibnamefont
  {Caviglia}}, \bibinfo {author} {\bibfnamefont {J.~M.}\ \bibnamefont
  {Triscone}}, \ and\ \bibinfo {author} {\bibfnamefont {A.}~\bibnamefont
  {Cavalleri}},\ }\bibfield  {title} {\enquote {\bibinfo {title} {Multiple
  supersonic phase fronts launched at a complex-oxide heterointerface},}\
  }\href {\doibase 10.1103/PhysRevLett.118.027401} {\bibfield  {journal}
  {\bibinfo  {journal} {Phys. Rev. Lett.}\ }\textbf {\bibinfo {volume} {118}},\
  \bibinfo {pages} {027401} (\bibinfo {year} {2017})},\ \Eprint
  {http://arxiv.org/abs/1612.04089} {1612.04089} \BibitemShut {NoStop}%
\bibitem [{\citenamefont {Wordenweber}(1999)}]{artc:Wordenweber1999}%
  \BibitemOpen
  \bibfield  {author} {\bibinfo {author} {\bibfnamefont {R.~W.}\ \bibnamefont
  {Wordenweber}},\ }\bibfield  {title} {\enquote {\bibinfo {title} {Growth of
  high-${T_c}$ thin films},}\ }\href
  {http://iopscience.iop.org/article/10.1088/0953-2048/12/6/202/pdf} {\bibfield
   {journal} {\bibinfo  {journal} {Supercond. Sci. Tech.}\ }\textbf {\bibinfo
  {volume} {12}},\ \bibinfo {pages} {79339--79344} (\bibinfo {year}
  {1999})}\BibitemShut {NoStop}%
\bibitem [{\citenamefont {Prellier}\ \emph {et~al.}(2001)\citenamefont
  {Prellier}, \citenamefont {Lecoeur},\ and\ \citenamefont
  {Mercey}}]{artc:Prellier2001}%
  \BibitemOpen
  \bibfield  {author} {\bibinfo {author} {\bibfnamefont {W.}~\bibnamefont
  {Prellier}}, \bibinfo {author} {\bibfnamefont {{\relax Ph}.}~\bibnamefont
  {Lecoeur}}, \ and\ \bibinfo {author} {\bibfnamefont {B.}~\bibnamefont
  {Mercey}},\ }\bibfield  {title} {\enquote {\bibinfo {title}
  {Colossal-magnetoresistive manganite thin films},}\ }\href {\doibase
  10.1088/0953-8984/13/48/201} {\bibfield  {journal} {\bibinfo  {journal}
  {Journal of Physics Condensed Matter}\ }\textbf {\bibinfo {volume} {13}},\
  \bibinfo {pages} {915--944} (\bibinfo {year} {2001})}\BibitemShut {NoStop}%
\bibitem [{\citenamefont {Middey}\ \emph {et~al.}(2016)\citenamefont {Middey},
  \citenamefont {Chakhalian}, \citenamefont {Mahadevan}, \citenamefont
  {Freeland}, \citenamefont {Millis},\ and\ \citenamefont
  {Sarma}}]{artc:Middey2016}%
  \BibitemOpen
  \bibfield  {author} {\bibinfo {author} {\bibfnamefont {S.}~\bibnamefont
  {Middey}}, \bibinfo {author} {\bibfnamefont {J.}~\bibnamefont {Chakhalian}},
  \bibinfo {author} {\bibfnamefont {P.}~\bibnamefont {Mahadevan}}, \bibinfo
  {author} {\bibfnamefont {J.}~\bibnamefont {Freeland}}, \bibinfo {author}
  {\bibfnamefont {A.}~\bibnamefont {Millis}}, \ and\ \bibinfo {author}
  {\bibfnamefont {D.}~\bibnamefont {Sarma}},\ }\bibfield  {title} {\enquote
  {\bibinfo {title} {Physics of ultrathin films and heterostructures of
  rare-earth nickelates},}\ }\href {\doibase
  10.1146/annurev-matsci-070115-032057} {\bibfield  {journal} {\bibinfo
  {journal} {Annual Review of Materials Research}\ }\textbf {\bibinfo {volume}
  {46}},\ \bibinfo {pages} {305--334} (\bibinfo {year} {2016})}\BibitemShut
  {NoStop}%
\bibitem [{\citenamefont {Sell}\ \emph {et~al.}(2008)\citenamefont {Sell},
  \citenamefont {Leitenstorfer},\ and\ \citenamefont {Huber}}]{artc:Sell2008}%
  \BibitemOpen
  \bibfield  {author} {\bibinfo {author} {\bibfnamefont {A.}~\bibnamefont
  {Sell}}, \bibinfo {author} {\bibfnamefont {A.}~\bibnamefont {Leitenstorfer}},
  \ and\ \bibinfo {author} {\bibfnamefont {R.}~\bibnamefont {Huber}},\
  }\bibfield  {title} {\enquote {\bibinfo {title} {Phase-locked generation and
  field-resolved detection of widely tunable terahertz pulses with amplitudes
  exceeding 100 {MV}/cm},}\ }\href {\doibase 10.1364/ol.33.002767} {\bibfield
  {journal} {\bibinfo  {journal} {Opt. Lett.}\ }\textbf {\bibinfo {volume}
  {33}},\ \bibinfo {pages} {2767--2769} (\bibinfo {year} {2008})}\BibitemShut
  {NoStop}%
\bibitem [{\citenamefont {Zhang}\ \emph {et~al.}(1994)\citenamefont {Zhang},
  \citenamefont {Choi}, \citenamefont {Flik},\ and\ \citenamefont
  {Anderson}}]{artc:Zhang1994}%
  \BibitemOpen
  \bibfield  {author} {\bibinfo {author} {\bibfnamefont {Z.~M.}\ \bibnamefont
  {Zhang}}, \bibinfo {author} {\bibfnamefont {B.~I.}\ \bibnamefont {Choi}},
  \bibinfo {author} {\bibfnamefont {M.~I.}\ \bibnamefont {Flik}}, \ and\
  \bibinfo {author} {\bibfnamefont {A.~C.}\ \bibnamefont {Anderson}},\
  }\bibfield  {title} {\enquote {\bibinfo {title} {Infrared refractive indices
  of {LaAlO$_3$}, {LaGaO$_3$}, and {NdGaO$_3$}},}\ }\href {\doibase
  10.1364/josab.11.002252} {\bibfield  {journal} {\bibinfo  {journal} {Journal
  of the Optical Society of America B}\ }\textbf {\bibinfo {volume} {11}},\
  \bibinfo {pages} {2252--2257} (\bibinfo {year} {1994})}\BibitemShut {NoStop}%
\bibitem [{\citenamefont {Abrashev}\ \emph {et~al.}(1999)\citenamefont
  {Abrashev}, \citenamefont {Litvinchuk}, \citenamefont {Iliev}, \citenamefont
  {Meng}, \citenamefont {Popov}, \citenamefont {Ivanov}, \citenamefont
  {Chakalov},\ and\ \citenamefont {Thomsen}}]{artc:Abrashev1999}%
  \BibitemOpen
  \bibfield  {author} {\bibinfo {author} {\bibfnamefont {M.~V.}\ \bibnamefont
  {Abrashev}}, \bibinfo {author} {\bibfnamefont {A.~P.}\ \bibnamefont
  {Litvinchuk}}, \bibinfo {author} {\bibfnamefont {M.~N.}\ \bibnamefont
  {Iliev}}, \bibinfo {author} {\bibfnamefont {R.~L.}\ \bibnamefont {Meng}},
  \bibinfo {author} {\bibfnamefont {V.~N.}\ \bibnamefont {Popov}}, \bibinfo
  {author} {\bibfnamefont {V.~G.}\ \bibnamefont {Ivanov}}, \bibinfo {author}
  {\bibfnamefont {R.~A.}\ \bibnamefont {Chakalov}}, \ and\ \bibinfo {author}
  {\bibfnamefont {C.}~\bibnamefont {Thomsen}},\ }\bibfield  {title} {\enquote
  {\bibinfo {title} {Comparative study of optical phonons in the rhombohedrally
  distorted perovskites {LaAlO$_3$} and {LaMnO$_3$}},}\ }\href {\doibase
  10.1103/PhysRevB.59.4146} {\bibfield  {journal} {\bibinfo  {journal} {Phys.
  Rev. B}\ }\textbf {\bibinfo {volume} {59}},\ \bibinfo {pages} {4146--4153}
  (\bibinfo {year} {1999})}\BibitemShut {NoStop}%
\bibitem [{\citenamefont {Lim}\ \emph {et~al.}(2002)\citenamefont {Lim},
  \citenamefont {Kriventsov}, \citenamefont {Jackson}, \citenamefont {Haeni},
  \citenamefont {Schlom}, \citenamefont {Balbashov}, \citenamefont {Uecker},
  \citenamefont {Reiche}, \citenamefont {Freeouf},\ and\ \citenamefont
  {Lucovsky}}]{artc:Lim2002}%
  \BibitemOpen
  \bibfield  {author} {\bibinfo {author} {\bibfnamefont {S.~G.}\ \bibnamefont
  {Lim}}, \bibinfo {author} {\bibfnamefont {S.}~\bibnamefont {Kriventsov}},
  \bibinfo {author} {\bibfnamefont {T.~N.}\ \bibnamefont {Jackson}}, \bibinfo
  {author} {\bibfnamefont {J.~H.}\ \bibnamefont {Haeni}}, \bibinfo {author}
  {\bibfnamefont {D.~G.}\ \bibnamefont {Schlom}}, \bibinfo {author}
  {\bibfnamefont {A.~M.}\ \bibnamefont {Balbashov}}, \bibinfo {author}
  {\bibfnamefont {R.}~\bibnamefont {Uecker}}, \bibinfo {author} {\bibfnamefont
  {P.}~\bibnamefont {Reiche}}, \bibinfo {author} {\bibfnamefont {J.~L.}\
  \bibnamefont {Freeouf}}, \ and\ \bibinfo {author} {\bibfnamefont
  {G.}~\bibnamefont {Lucovsky}},\ }\bibfield  {title} {\enquote {\bibinfo
  {title} {Dielectric functions and optical bandgaps of high-{K} dielectrics
  for metal-oxide-semiconductor field-effect transistors by far ultraviolet
  spectroscopic ellipsometry},}\ }\href {\doibase 10.1063/1.1456246} {\bibfield
   {journal} {\bibinfo  {journal} {Journal of Applied Physics}\ }\textbf
  {\bibinfo {volume} {91}},\ \bibinfo {pages} {4500--4505} (\bibinfo {year}
  {2002})}\BibitemShut {NoStop}%
\bibitem [{\citenamefont {Chernova}\ \emph {et~al.}(2017)\citenamefont
  {Chernova}, \citenamefont {Brooks}, \citenamefont {Chvostova}, \citenamefont
  {Bryknar}, \citenamefont {Dejneka},\ and\ \citenamefont
  {Tyunina}}]{artc:Chernova2017}%
  \BibitemOpen
  \bibfield  {author} {\bibinfo {author} {\bibfnamefont {E.}~\bibnamefont
  {Chernova}}, \bibinfo {author} {\bibfnamefont {C.}~\bibnamefont {Brooks}},
  \bibinfo {author} {\bibfnamefont {D.}~\bibnamefont {Chvostova}}, \bibinfo
  {author} {\bibfnamefont {Z.}~\bibnamefont {Bryknar}}, \bibinfo {author}
  {\bibfnamefont {A.}~\bibnamefont {Dejneka}}, \ and\ \bibinfo {author}
  {\bibfnamefont {M.}~\bibnamefont {Tyunina}},\ }\bibfield  {title} {\enquote
  {\bibinfo {title} {Optical {NIR-VIS-VUV} constants of advanced substrates for
  thin-film devices},}\ }\href {\doibase 10.1364/ome.7.003844} {\bibfield
  {journal} {\bibinfo  {journal} {Opt. Mater. Express}\ }\textbf {\bibinfo
  {volume} {7}},\ \bibinfo {pages} {3844--3862} (\bibinfo {year} {2017})},\
  \Eprint {http://arxiv.org/abs/1708.06979} {1708.06979} \BibitemShut {NoStop}%
\bibitem [{\citenamefont {Scott}(1969)}]{artc:Scott1969}%
  \BibitemOpen
  \bibfield  {author} {\bibinfo {author} {\bibfnamefont {J.~F.}\ \bibnamefont
  {Scott}},\ }\bibfield  {title} {\enquote {\bibinfo {title} {Raman study of
  trigonal-cubic phase transitions in rare-earth aluminates},}\ }\href
  {\doibase 10.1103/PhysRev.183.823} {\bibfield  {journal} {\bibinfo  {journal}
  {Phys. Rev.}\ }\textbf {\bibinfo {volume} {183}},\ \bibinfo {pages}
  {823--825} (\bibinfo {year} {1969})}\BibitemShut {NoStop}%
\bibitem [{\citenamefont {Matsuda}\ \emph {et~al.}(2004)\citenamefont
  {Matsuda}, \citenamefont {Wright}, \citenamefont {Hurley}, \citenamefont
  {Gusev},\ and\ \citenamefont {Shimizu}}]{artc:Matsuda2004}%
  \BibitemOpen
  \bibfield  {author} {\bibinfo {author} {\bibfnamefont {O.}~\bibnamefont
  {Matsuda}}, \bibinfo {author} {\bibfnamefont {O.~B.}\ \bibnamefont {Wright}},
  \bibinfo {author} {\bibfnamefont {D.~H.}\ \bibnamefont {Hurley}}, \bibinfo
  {author} {\bibfnamefont {V.~E.}\ \bibnamefont {Gusev}}, \ and\ \bibinfo
  {author} {\bibfnamefont {K.}~\bibnamefont {Shimizu}},\ }\bibfield  {title}
  {\enquote {\bibinfo {title} {Coherent shear phonon generation and detection
  with ultrashort optical pulses},}\ }\href {\doibase
  10.1103/PhysRevLett.93.095501} {\bibfield  {journal} {\bibinfo  {journal}
  {Physical Review Letters}\ }\textbf {\bibinfo {volume} {93}},\ \bibinfo
  {pages} {095501} (\bibinfo {year} {2004})}\BibitemShut {NoStop}%
\bibitem [{\citenamefont {Carpenter}\ \emph {et~al.}(2010)\citenamefont
  {Carpenter}, \citenamefont {Sinogeikin}, \citenamefont {Bass}, \citenamefont
  {Lakshtanov},\ and\ \citenamefont {Jacobsen}}]{artc:Carpenter2010}%
  \BibitemOpen
  \bibfield  {author} {\bibinfo {author} {\bibfnamefont {M.~A.}\ \bibnamefont
  {Carpenter}}, \bibinfo {author} {\bibfnamefont {S.~V.}\ \bibnamefont
  {Sinogeikin}}, \bibinfo {author} {\bibfnamefont {J.~D.}\ \bibnamefont
  {Bass}}, \bibinfo {author} {\bibfnamefont {D.~L.}\ \bibnamefont
  {Lakshtanov}}, \ and\ \bibinfo {author} {\bibfnamefont {S.~D.}\ \bibnamefont
  {Jacobsen}},\ }\bibfield  {title} {\enquote {\bibinfo {title} {Elastic
  relaxations associated with the {Pm$\bar{3}$m-R$\bar{3}$c} transition in
  {LaAlO$_3$}: I. single crystal elastic moduli at room temperature},}\ }\href
  {\doibase 10.1088/0953-8984/22/3/035403} {\bibfield  {journal} {\bibinfo
  {journal} {J. Condens. Matter. Phys.}\ }\textbf {\bibinfo {volume} {22}}
  (\bibinfo {year} {2010}),\ 10.1088/0953-8984/22/3/035403}\BibitemShut
  {NoStop}%
\bibitem [{\citenamefont {Cancellieri}\ \emph {et~al.}(2011)\citenamefont
  {Cancellieri}, \citenamefont {Fontaine}, \citenamefont {Gariglio},
  \citenamefont {Reyren}, \citenamefont {Caviglia}, \citenamefont {F\^{e}te},
  \citenamefont {Leake}, \citenamefont {Pauli}, \citenamefont {Willmott},
  \citenamefont {Stengel}, \citenamefont {Ghosez},\ and\ \citenamefont
  {Triscone}}]{artc:Cancellieri2011}%
  \BibitemOpen
  \bibfield  {author} {\bibinfo {author} {\bibfnamefont {C.}~\bibnamefont
  {Cancellieri}}, \bibinfo {author} {\bibfnamefont {D.}~\bibnamefont
  {Fontaine}}, \bibinfo {author} {\bibfnamefont {S.}~\bibnamefont {Gariglio}},
  \bibinfo {author} {\bibfnamefont {N.}~\bibnamefont {Reyren}}, \bibinfo
  {author} {\bibfnamefont {A.~D.}\ \bibnamefont {Caviglia}}, \bibinfo {author}
  {\bibfnamefont {A.}~\bibnamefont {F\^{e}te}}, \bibinfo {author}
  {\bibfnamefont {S.~J.}\ \bibnamefont {Leake}}, \bibinfo {author}
  {\bibfnamefont {S.~A.}\ \bibnamefont {Pauli}}, \bibinfo {author}
  {\bibfnamefont {P.~R.}\ \bibnamefont {Willmott}}, \bibinfo {author}
  {\bibfnamefont {M.}~\bibnamefont {Stengel}}, \bibinfo {author} {\bibfnamefont
  {P.}~\bibnamefont {Ghosez}}, \ and\ \bibinfo {author} {\bibfnamefont {J.~M.}\
  \bibnamefont {Triscone}},\ }\bibfield  {title} {\enquote {\bibinfo {title}
  {Electrostriction at the {LaAlO$_3$/SrTiO$_3$} interface},}\ }\href {\doibase
  10.1103/PhysRevLett.107.056102} {\bibfield  {journal} {\bibinfo  {journal}
  {Physical Review Letters}\ }\textbf {\bibinfo {volume} {107}},\ \bibinfo
  {pages} {056102} (\bibinfo {year} {2011})}\BibitemShut {NoStop}%
\bibitem [{\citenamefont {Ruello}\ and\ \citenamefont
  {Gusev}(2015)}]{artc:Ruello2015}%
  \BibitemOpen
  \bibfield  {author} {\bibinfo {author} {\bibfnamefont {P.}~\bibnamefont
  {Ruello}}\ and\ \bibinfo {author} {\bibfnamefont {V.~E.}\ \bibnamefont
  {Gusev}},\ }\bibfield  {title} {\enquote {\bibinfo {title} {Physical
  mechanisms of coherent acoustic phonons generation by ultrafast laser
  action},}\ }\href {\doibase 10.1016/j.ultras.2014.06.004} {\bibfield
  {journal} {\bibinfo  {journal} {Ultrasonics}\ }\textbf {\bibinfo {volume}
  {56}},\ \bibinfo {pages} {21--35} (\bibinfo {year} {2015})}\BibitemShut
  {NoStop}%
\bibitem [{\citenamefont {F{\"{o}}rst}\ \emph {et~al.}(2011)\citenamefont
  {F{\"{o}}rst}, \citenamefont {Manzoni}, \citenamefont {Kaiser}, \citenamefont
  {Tomioka}, \citenamefont {Tokura}, \citenamefont {Merlin},\ and\
  \citenamefont {Cavalleri}}]{artc:Forst2011_2}%
  \BibitemOpen
  \bibfield  {author} {\bibinfo {author} {\bibfnamefont {M.}~\bibnamefont
  {F{\"{o}}rst}}, \bibinfo {author} {\bibfnamefont {C.}~\bibnamefont
  {Manzoni}}, \bibinfo {author} {\bibfnamefont {S.}~\bibnamefont {Kaiser}},
  \bibinfo {author} {\bibfnamefont {Y.}~\bibnamefont {Tomioka}}, \bibinfo
  {author} {\bibfnamefont {Y.}~\bibnamefont {Tokura}}, \bibinfo {author}
  {\bibfnamefont {R.}~\bibnamefont {Merlin}}, \ and\ \bibinfo {author}
  {\bibfnamefont {A.}~\bibnamefont {Cavalleri}},\ }\bibfield  {title} {\enquote
  {\bibinfo {title} {Nonlinear phononics as an ultrafast route to lattice
  control},}\ }\href {\doibase 10.1038/nphys2055} {\bibfield  {journal}
  {\bibinfo  {journal} {Nat. Phys.}\ }\textbf {\bibinfo {volume} {7}},\
  \bibinfo {pages} {854--856} (\bibinfo {year} {2011})}\BibitemShut {NoStop}%
\bibitem [{\citenamefont {et~al.}(2020)}]{artc:gonze2020}%
  \BibitemOpen
  \bibfield  {author} {\bibinfo {author} {\bibfnamefont {X.~G.}\ \bibnamefont
  {et~al.}},\ }\bibfield  {title} {\enquote {\bibinfo {title} {The
  {A}binitproject: {I}mpact, environment and recent developments},}\ }\href
  {\doibase https://doi.org/10.1016/j.cpc.2019.107042} {\bibfield  {journal}
  {\bibinfo  {journal} {Comput. Phys. Commun.}\ }\textbf {\bibinfo {volume}
  {248}},\ \bibinfo {pages} {107042} (\bibinfo {year} {2020})}\BibitemShut
  {NoStop}%
\bibitem [{\citenamefont {Afanasiev}\ \emph {et~al.}(2014)\citenamefont
  {Afanasiev}, \citenamefont {Razdolski}, \citenamefont {Skibinsky},
  \citenamefont {Bolotin}, \citenamefont {Yagupov}, \citenamefont {Strugatsky},
  \citenamefont {Kirilyuk}, \citenamefont {Rasing},\ and\ \citenamefont
  {Kimel}}]{artc:Afanasiev2014}%
  \BibitemOpen
  \bibfield  {author} {\bibinfo {author} {\bibfnamefont {D.}~\bibnamefont
  {Afanasiev}}, \bibinfo {author} {\bibfnamefont {I.}~\bibnamefont
  {Razdolski}}, \bibinfo {author} {\bibfnamefont {K.~M.}\ \bibnamefont
  {Skibinsky}}, \bibinfo {author} {\bibfnamefont {D.}~\bibnamefont {Bolotin}},
  \bibinfo {author} {\bibfnamefont {S.~V.}\ \bibnamefont {Yagupov}}, \bibinfo
  {author} {\bibfnamefont {M.~B.}\ \bibnamefont {Strugatsky}}, \bibinfo
  {author} {\bibfnamefont {A.}~\bibnamefont {Kirilyuk}}, \bibinfo {author}
  {\bibfnamefont {{\relax Th}.}~\bibnamefont {Rasing}}, \ and\ \bibinfo
  {author} {\bibfnamefont {A.~V.}\ \bibnamefont {Kimel}},\ }\bibfield  {title}
  {\enquote {\bibinfo {title} {Laser excitation of lattice-driven anharmonic
  magnetization dynamics in dielectric {FeBO$_3$}},}\ }\href {\doibase
  10.1103/PhysRevLett.112.147403} {\bibfield  {journal} {\bibinfo  {journal}
  {Phys. Rev. Lett.}\ }\textbf {\bibinfo {volume} {112}},\ \bibinfo {pages}
  {147403} (\bibinfo {year} {2014})}\BibitemShut {NoStop}%
\bibitem [{\citenamefont {Zubko}\ \emph {et~al.}(2013)\citenamefont {Zubko},
  \citenamefont {Catalan},\ and\ \citenamefont {Tagantsev}}]{artc:Zubko2013}%
  \BibitemOpen
  \bibfield  {author} {\bibinfo {author} {\bibfnamefont {P.}~\bibnamefont
  {Zubko}}, \bibinfo {author} {\bibfnamefont {G.}~\bibnamefont {Catalan}}, \
  and\ \bibinfo {author} {\bibfnamefont {A.~K.}\ \bibnamefont {Tagantsev}},\
  }\bibfield  {title} {\enquote {\bibinfo {title} {Flexoelectric effect in
  solids},}\ }\href {\doibase 10.1146/annurev-matsci-071312-121634} {\bibfield
  {journal} {\bibinfo  {journal} {Annual Review of Materials Research}\
  }\textbf {\bibinfo {volume} {43}},\ \bibinfo {pages} {387--421} (\bibinfo
  {year} {2013})}\BibitemShut {NoStop}%
\bibitem [{\citenamefont {Kohmoto}\ \emph {et~al.}(2011)\citenamefont
  {Kohmoto}, \citenamefont {Masui}, \citenamefont {Abe}, \citenamefont
  {Moriyasu},\ and\ \citenamefont {Tanaka}}]{artc:Kohmoto2011}%
  \BibitemOpen
  \bibfield  {author} {\bibinfo {author} {\bibfnamefont {T.}~\bibnamefont
  {Kohmoto}}, \bibinfo {author} {\bibfnamefont {M.}~\bibnamefont {Masui}},
  \bibinfo {author} {\bibfnamefont {M.}~\bibnamefont {Abe}}, \bibinfo {author}
  {\bibfnamefont {T.}~\bibnamefont {Moriyasu}}, \ and\ \bibinfo {author}
  {\bibfnamefont {K.}~\bibnamefont {Tanaka}},\ }\bibfield  {title} {\enquote
  {\bibinfo {title} {Ultrafast dynamics of soft phonon modes in perovskite
  dielectrics observed by coherent phonon spectroscopy},}\ }\href {\doibase
  10.1103/PhysRevB.83.064304} {\bibfield  {journal} {\bibinfo  {journal} {Phys.
  Rev. B}\ }\textbf {\bibinfo {volume} {83}} (\bibinfo {year} {2011}),\
  10.1103/PhysRevB.83.064304}\BibitemShut {NoStop}%
\bibitem [{\citenamefont {Klemens}(1966)}]{artc:Klemens1966}%
  \BibitemOpen
  \bibfield  {author} {\bibinfo {author} {\bibfnamefont {P.~G.}\ \bibnamefont
  {Klemens}},\ }\bibfield  {title} {\enquote {\bibinfo {title} {Anharmonic
  decay of optical phonon in diamond},}\ }\href {\doibase
  10.1103/PhysRevB.11.3206} {\bibfield  {journal} {\bibinfo  {journal} {Phys.
  Rev. B}\ }\textbf {\bibinfo {volume} {138}},\ \bibinfo {pages} {845--848}
  (\bibinfo {year} {1966})}\BibitemShut {NoStop}%
\bibitem [{\citenamefont {Hohenberg}\ and\ \citenamefont
  {Kohn}(1964)}]{Hohenberg1964}%
  \BibitemOpen
  \bibfield  {author} {\bibinfo {author} {\bibfnamefont {P.}~\bibnamefont
  {Hohenberg}}\ and\ \bibinfo {author} {\bibfnamefont {W.}~\bibnamefont
  {Kohn}},\ }\bibfield  {title} {\enquote {\bibinfo {title} {Inhomogeneous
  electron gas},}\ }\href {\doibase 10.1103/PhysRev.136.B864} {\bibfield
  {journal} {\bibinfo  {journal} {Phys. Rev.}\ }\textbf {\bibinfo {volume}
  {136}},\ \bibinfo {pages} {B864} (\bibinfo {year} {1964})}\BibitemShut
  {NoStop}%
\bibitem [{\citenamefont {Kohn}\ and\ \citenamefont {Sham}(1965)}]{Kohn1965}%
  \BibitemOpen
  \bibfield  {author} {\bibinfo {author} {\bibfnamefont {W.}~\bibnamefont
  {Kohn}}\ and\ \bibinfo {author} {\bibfnamefont {L.~J.}\ \bibnamefont
  {Sham}},\ }\bibfield  {title} {\enquote {\bibinfo {title} {Self-consistent
  equations including exchange and correlation effects},}\ }\href {\doibase
  10.1103/PhysRev.140.A1133} {\bibfield  {journal} {\bibinfo  {journal} {Phys.
  Rev.}\ }\textbf {\bibinfo {volume} {140}},\ \bibinfo {pages} {A1133}
  (\bibinfo {year} {1965})}\BibitemShut {NoStop}%
\bibitem [{\citenamefont {Gonze}\ \emph {et~al.}(2002)\citenamefont {Gonze},
  \citenamefont {Beuken}, \citenamefont {Caracas}, \citenamefont {Detraux},
  \citenamefont {Fuchs}, \citenamefont {Rignanese}, \citenamefont {Sindic},
  \citenamefont {Verstraete}, \citenamefont {Zerah}, \citenamefont {Jollet},
  \citenamefont {Torrent}, \citenamefont {Roy}, \citenamefont {Mikami},
  \citenamefont {Ghosez}, \citenamefont {Raty},\ and\ \citenamefont
  {Allan}}]{Gonze2002}%
  \BibitemOpen
  \bibfield  {author} {\bibinfo {author} {\bibfnamefont {X.}~\bibnamefont
  {Gonze}}, \bibinfo {author} {\bibfnamefont {J.-M.}\ \bibnamefont {Beuken}},
  \bibinfo {author} {\bibfnamefont {R.}~\bibnamefont {Caracas}}, \bibinfo
  {author} {\bibfnamefont {F.}~\bibnamefont {Detraux}}, \bibinfo {author}
  {\bibfnamefont {M.}~\bibnamefont {Fuchs}}, \bibinfo {author} {\bibfnamefont
  {G.-M.}\ \bibnamefont {Rignanese}}, \bibinfo {author} {\bibfnamefont
  {L.}~\bibnamefont {Sindic}}, \bibinfo {author} {\bibfnamefont
  {M.}~\bibnamefont {Verstraete}}, \bibinfo {author} {\bibfnamefont
  {G.}~\bibnamefont {Zerah}}, \bibinfo {author} {\bibfnamefont
  {F.}~\bibnamefont {Jollet}}, \bibinfo {author} {\bibfnamefont
  {M.}~\bibnamefont {Torrent}}, \bibinfo {author} {\bibfnamefont
  {A.}~\bibnamefont {Roy}}, \bibinfo {author} {\bibfnamefont {M.}~\bibnamefont
  {Mikami}}, \bibinfo {author} {\bibfnamefont {P.}~\bibnamefont {Ghosez}},
  \bibinfo {author} {\bibfnamefont {J.-Y.}\ \bibnamefont {Raty}}, \ and\
  \bibinfo {author} {\bibfnamefont {D.}~\bibnamefont {Allan}},\ }\bibfield
  {title} {\enquote {\bibinfo {title} {First-principles computation of material
  properties: The {ABINIT} software project},}\ }\href {\doibase
  10.1016/S0927-0256(02)00325-7} {\bibfield  {journal} {\bibinfo  {journal}
  {Comput. Mater. Sci.}\ }\textbf {\bibinfo {volume} {25}},\ \bibinfo {pages}
  {478--492} (\bibinfo {year} {2002})}\BibitemShut {NoStop}%
\bibitem [{\citenamefont {Torrent}\ \emph {et~al.}(2008)\citenamefont
  {Torrent}, \citenamefont {Jollet}, \citenamefont {Bottin}, \citenamefont
  {Zerah},\ and\ \citenamefont {Gonze}}]{Torrent2008}%
  \BibitemOpen
  \bibfield  {author} {\bibinfo {author} {\bibfnamefont {M.}~\bibnamefont
  {Torrent}}, \bibinfo {author} {\bibfnamefont {F.}~\bibnamefont {Jollet}},
  \bibinfo {author} {\bibfnamefont {F.}~\bibnamefont {Bottin}}, \bibinfo
  {author} {\bibfnamefont {G.}~\bibnamefont {Zerah}}, \ and\ \bibinfo {author}
  {\bibfnamefont {X.}~\bibnamefont {Gonze}},\ }\bibfield  {title} {\enquote
  {\bibinfo {title} {Implementation of the projector augmented-wave method in
  the {ABINIT} code: Application to the study of iron under pressure},}\ }\href
  {\doibase 10.1016/j.commatsci.2007.07.020} {\bibfield  {journal} {\bibinfo
  {journal} {Computational Materials Science}\ }\textbf {\bibinfo {volume}
  {42}},\ \bibinfo {pages} {337--351} (\bibinfo {year} {2008})}\BibitemShut
  {NoStop}%
\bibitem [{\citenamefont {Hamann}(2013)}]{Hamann_2013}%
  \BibitemOpen
  \bibfield  {author} {\bibinfo {author} {\bibfnamefont {D.~R.}\ \bibnamefont
  {Hamann}},\ }\bibfield  {title} {\enquote {\bibinfo {title} {Optimized
  norm-conserving {V}anderbilt pseudopotentials},}\ }\href {\doibase
  10.1103/PhysRevB.88.085117} {\bibfield  {journal} {\bibinfo  {journal} {Phys.
  Rev. B}\ }\textbf {\bibinfo {volume} {88}},\ \bibinfo {pages} {085117}
  (\bibinfo {year} {2013})}\BibitemShut {NoStop}%
\bibitem [{\citenamefont {van Setten}\ \emph {et~al.}(2018)\citenamefont {van
  Setten}, \citenamefont {Giantomassi}, \citenamefont {Bousquet}, \citenamefont
  {Verstraete}, \citenamefont {Hamann}, \citenamefont {Gonze},\ and\
  \citenamefont {Rignanese}}]{Pspweb}%
  \BibitemOpen
  \bibfield  {author} {\bibinfo {author} {\bibfnamefont {M.}~\bibnamefont {van
  Setten}}, \bibinfo {author} {\bibfnamefont {M.}~\bibnamefont {Giantomassi}},
  \bibinfo {author} {\bibfnamefont {E.}~\bibnamefont {Bousquet}}, \bibinfo
  {author} {\bibfnamefont {M.}~\bibnamefont {Verstraete}}, \bibinfo {author}
  {\bibfnamefont {D.}~\bibnamefont {Hamann}}, \bibinfo {author} {\bibfnamefont
  {X.}~\bibnamefont {Gonze}}, \ and\ \bibinfo {author} {\bibfnamefont {G.-M.}\
  \bibnamefont {Rignanese}},\ }\bibfield  {title} {\enquote {\bibinfo {title}
  {The pseudodojo: Training and grading a 85 element optimized norm-conserving
  pseudopotential table},}\ }\href {\doibase
  https://doi.org/10.1016/j.cpc.2018.01.012} {\bibfield  {journal} {\bibinfo
  {journal} {Comput. Phys. Commun.}\ }\textbf {\bibinfo {volume} {226}},\
  \bibinfo {pages} {39 -- 54} (\bibinfo {year} {2018})}\BibitemShut {NoStop}%
\bibitem [{\citenamefont {P~Perdew}\ \emph {et~al.}(2008)\citenamefont
  {P~Perdew}, \citenamefont {Ruzsinszky}, \citenamefont {Csonka}, \citenamefont
  {A~Vydrov}, \citenamefont {E~Scuseria}, \citenamefont {Constantin},
  \citenamefont {Zhou},\ and\ \citenamefont {Burke}}]{Perdew2008}%
  \BibitemOpen
  \bibfield  {author} {\bibinfo {author} {\bibfnamefont {J.}~\bibnamefont
  {P~Perdew}}, \bibinfo {author} {\bibfnamefont {A.}~\bibnamefont
  {Ruzsinszky}}, \bibinfo {author} {\bibfnamefont {G.}~\bibnamefont {Csonka}},
  \bibinfo {author} {\bibfnamefont {O.}~\bibnamefont {A~Vydrov}}, \bibinfo
  {author} {\bibfnamefont {G.}~\bibnamefont {E~Scuseria}}, \bibinfo {author}
  {\bibfnamefont {L.}~\bibnamefont {Constantin}}, \bibinfo {author}
  {\bibfnamefont {X.}~\bibnamefont {Zhou}}, \ and\ \bibinfo {author}
  {\bibfnamefont {K.}~\bibnamefont {Burke}},\ }\bibfield  {title} {\enquote
  {\bibinfo {title} {Restoring the density-gradient expansion for exchange in
  solids and surfaces},}\ }\href {\doibase 10.1103/PhysRevLett.100.136406}
  {\bibfield  {journal} {\bibinfo  {journal} {Phys. Rev. Lett.}\ }\textbf
  {\bibinfo {volume} {100}},\ \bibinfo {pages} {136406} (\bibinfo {year}
  {2008})}\BibitemShut {NoStop}%
\bibitem [{\citenamefont {Gonze}\ and\ \citenamefont {Lee}(1997)}]{DFPT-Gonze}%
  \BibitemOpen
  \bibfield  {author} {\bibinfo {author} {\bibfnamefont {X.}~\bibnamefont
  {Gonze}}\ and\ \bibinfo {author} {\bibfnamefont {C.}~\bibnamefont {Lee}},\
  }\bibfield  {title} {\enquote {\bibinfo {title} {Dynamical matrices, born
  effective charges, dielectric permittivity tensors, and interatomic force
  constants from density-functional perturbation theory},}\ }\href {\doibase
  10.1103/PhysRevB.55.10355} {\bibfield  {journal} {\bibinfo  {journal} {Phys.
  Rev. B}\ }\textbf {\bibinfo {volume} {55}},\ \bibinfo {pages} {10355--10368}
  (\bibinfo {year} {1997})}\BibitemShut {NoStop}%
\bibitem [{\citenamefont {Gonze}(1997)}]{DFPT-Gonzea}%
  \BibitemOpen
  \bibfield  {author} {\bibinfo {author} {\bibfnamefont {X.}~\bibnamefont
  {Gonze}},\ }\bibfield  {title} {\enquote {\bibinfo {title} {First-principles
  responses of solids to atomic displacements and homogeneous electric fields:
  Implementation of a conjugate-gradient algorithm},}\ }\href {\doibase
  10.1103/PhysRevB.55.10337} {\bibfield  {journal} {\bibinfo  {journal} {Phys.
  Rev. B}\ }\textbf {\bibinfo {volume} {55}},\ \bibinfo {pages} {10337--10354}
  (\bibinfo {year} {1997})}\BibitemShut {NoStop}%
\bibitem [{\citenamefont {Fredrickson}\ \emph {et~al.}(2016)\citenamefont
  {Fredrickson}, \citenamefont {Lin}, \citenamefont {Zollner},\ and\
  \citenamefont {Demkov}}]{Fredrickson-2016}%
  \BibitemOpen
  \bibfield  {author} {\bibinfo {author} {\bibfnamefont {K.~D.}\ \bibnamefont
  {Fredrickson}}, \bibinfo {author} {\bibfnamefont {C.}~\bibnamefont {Lin}},
  \bibinfo {author} {\bibfnamefont {S.}~\bibnamefont {Zollner}}, \ and\
  \bibinfo {author} {\bibfnamefont {A.~A.}\ \bibnamefont {Demkov}},\ }\bibfield
   {title} {\enquote {\bibinfo {title} {Theoretical study of negative optical
  mode splitting in},}\ }\href {\doibase 10.1103/PhysRevB.93.134301} {\bibfield
   {journal} {\bibinfo  {journal} {Phys. Rev. B}\ }\textbf {\bibinfo {volume}
  {93}},\ \bibinfo {pages} {134301} (\bibinfo {year} {2016})}\BibitemShut
  {NoStop}%
\bibitem [{\citenamefont {Hatt}\ and\ \citenamefont
  {Spaldin}(2010)}]{Hatt2010}%
  \BibitemOpen
  \bibfield  {author} {\bibinfo {author} {\bibfnamefont {A.~J.}\ \bibnamefont
  {Hatt}}\ and\ \bibinfo {author} {\bibfnamefont {N.~A.}\ \bibnamefont
  {Spaldin}},\ }\bibfield  {title} {\enquote {\bibinfo {title} {Structural
  phases of strained},}\ }\href {\doibase 10.1103/PhysRevB.82.195402}
  {\bibfield  {journal} {\bibinfo  {journal} {Phys. Rev. B}\ }\textbf {\bibinfo
  {volume} {82}},\ \bibinfo {pages} {195402} (\bibinfo {year}
  {2010})}\BibitemShut {NoStop}%
\bibitem [{\citenamefont {Geller}\ and\ \citenamefont
  {Bala}(1956)}]{Geller1956}%
  \BibitemOpen
  \bibfield  {author} {\bibinfo {author} {\bibfnamefont {S.}~\bibnamefont
  {Geller}}\ and\ \bibinfo {author} {\bibfnamefont {V.~B.}\ \bibnamefont
  {Bala}},\ }\bibfield  {title} {\enquote {\bibinfo {title} {Crystallographic
  studies of perovskite-like compounds. ii. rare earth alluminates},}\ }\href
  {\doibase https://doi.org/10.1107/S0365110X56002965} {\bibfield  {journal}
  {\bibinfo  {journal} {Acta Cryst.}\ }\textbf {\bibinfo {volume} {9}},\
  \bibinfo {pages} {1019} (\bibinfo {year} {1956})}\BibitemShut {NoStop}%
\bibitem [{\citenamefont {M\"uller}\ \emph {et~al.}(1968)\citenamefont
  {M\"uller}, \citenamefont {Berlinger},\ and\ \citenamefont
  {Waldner}}]{Muller1968}%
  \BibitemOpen
  \bibfield  {author} {\bibinfo {author} {\bibfnamefont {K.~A.}\ \bibnamefont
  {M\"uller}}, \bibinfo {author} {\bibfnamefont {W.}~\bibnamefont {Berlinger}},
  \ and\ \bibinfo {author} {\bibfnamefont {F.}~\bibnamefont {Waldner}},\
  }\bibfield  {title} {\enquote {\bibinfo {title} {Characteristic structural
  phase transition in perovskite-type compounds},}\ }\href {\doibase
  10.1103/PhysRevLett.21.814} {\bibfield  {journal} {\bibinfo  {journal} {Phys.
  Rev. Lett.}\ }\textbf {\bibinfo {volume} {21}},\ \bibinfo {pages} {814--817}
  (\bibinfo {year} {1968})}\BibitemShut {NoStop}%
\bibitem [{\citenamefont {Delugas}\ \emph {et~al.}(2005)\citenamefont
  {Delugas}, \citenamefont {Fiorentini},\ and\ \citenamefont
  {Filippetti}}]{Delugas_2005}%
  \BibitemOpen
  \bibfield  {author} {\bibinfo {author} {\bibfnamefont {P.}~\bibnamefont
  {Delugas}}, \bibinfo {author} {\bibfnamefont {V.}~\bibnamefont {Fiorentini}},
  \ and\ \bibinfo {author} {\bibfnamefont {A.}~\bibnamefont {Filippetti}},\
  }\bibfield  {title} {\enquote {\bibinfo {title} {Dielectric properties and
  long-wavelength optical modes of the high-$\kappa$ oxide {LaAlO$_3$}},}\
  }\href {\doibase 10.1103/PhysRevB.71.134302} {\bibfield  {journal} {\bibinfo
  {journal} {Phys. Rev. B}\ }\textbf {\bibinfo {volume} {71}},\ \bibinfo
  {pages} {134302} (\bibinfo {year} {2005})}\BibitemShut {NoStop}%
\bibitem [{\citenamefont {Calvani}\ \emph {et~al.}(1991)\citenamefont
  {Calvani}, \citenamefont {Capizzi}, \citenamefont {Donato}, \citenamefont
  {Dore}, \citenamefont {Lupi}, \citenamefont {Maselli},\ and\ \citenamefont
  {Varsamis}}]{Calvani_1991}%
  \BibitemOpen
  \bibfield  {author} {\bibinfo {author} {\bibfnamefont {P.}~\bibnamefont
  {Calvani}}, \bibinfo {author} {\bibfnamefont {M.}~\bibnamefont {Capizzi}},
  \bibinfo {author} {\bibfnamefont {F.}~\bibnamefont {Donato}}, \bibinfo
  {author} {\bibfnamefont {P.}~\bibnamefont {Dore}}, \bibinfo {author}
  {\bibfnamefont {S.}~\bibnamefont {Lupi}}, \bibinfo {author} {\bibfnamefont
  {P.}~\bibnamefont {Maselli}}, \ and\ \bibinfo {author} {\bibfnamefont
  {C.}~\bibnamefont {Varsamis}},\ }\bibfield  {title} {\enquote {\bibinfo
  {title} {Infrared optical properties of perovskite substrates for
  high-{T$_\text{c}$} superconducting films},}\ }\href {\doibase
  https://doi.org/10.1016/0921-4534(91)90113-D} {\bibfield  {journal} {\bibinfo
   {journal} {Physica {C} Supercond.}\ }\textbf {\bibinfo {volume} {181}},\
  \bibinfo {pages} {289--295} (\bibinfo {year} {1991})}\BibitemShut {NoStop}%
\end{thebibliography}%

\end{document}